\documentclass{ws-ijgmmp}

\usepackage{xcolor}
\usepackage[verbose,hypertexnames=false]{hyperref}
\hypersetup{colorlinks=false,allbordercolors=blue,pdfborderstyle={/S/U/W 1}}

\begin{document}

\title{Primary Constraints of Newer General Relativity}

\author{Carmen Ferrara}

\address{Scuola Superiore Meridionale, Largo San Marcellino 10, I-80138 Napoli, Italy,\\
Istituto Nazionale di Fisica Nucleare, Sezione di Napoli,
Complesso Universitario di Monte S. Angelo, Via Cinthia Edificio 6, I-80126 Napoli, Italy
\\
\email{carmen.ferrara-ssm@unina.it} }

\author{Alexey Golovnev}

\address{Centre for Theoretical Physics, the British University in Egypt,
BUE 11837, El Sherouk City, Cairo Governorate, Egypt\\
agolovnev@yandex.ru }

\author{Mar\'ia Jos\'e Guzm\'an}

\address{Laboratory of Theoretical Physics, Institute of Physics, University of Tartu, W. Ostwaldi 1, 50411 Tartu, Estonia\\
mjguzman@ut.ee }

\maketitle

%\begin{history}
%\received{(Day Month Year)}
%\revised{(Day Month Year)}
%\accepted{(Day Month Year)}
%\published{(Day Month Year)}
%\end{history}

\begin{abstract}
We study the primary constraint structure of Newer General Relativity, a gravity theory based on a torsionless teleparallel geometry. The gravitational action is built from a scalar formed by quadratic combinations of the nonmetricity tensor, with arbitrary coefficients $c_i$ in the Lagrangian. We decompose the Lagrangian and compute the canonical momenta conjugate to the metric. We characterize the primary constraints arising from these momenta by identifying when the map between velocities and momenta becomes non-invertible, and organize the outcome through a fully nonlinear decomposition into scalar, vector and tensor sectors. Comparing with previous results in the literature, we recover five and three primary constraints associated with the tensor and vector sectors, respectively. We also identify a previously unreported degeneracy in the scalar sector, which yields either one or two scalar primary constraints depending on the conditions imposed on the parameters $c_i$. Finally, we obtain the primary constraints associated with the covariant formulation of symmetric teleparallel gravity.
\end{abstract}



\section{Introduction}

General relativity (GR) remains the most accurate classical description of gravitation available, having passed an extensive battery of tests ranging from precision solar system measurements to strong-field probes and gravitational-wave observations \cite{Will1993,Ni_2016,De_Marchi_2020,Abbott2017-NS,LIGOScientific2021psn,Abbott2022,Stairs_2003,Kramer_2021}. Yet, despite this empirical success, there are well-known theoretical and observational reasons to explore extensions of GR, and more broadly, alternative geometric formulations of gravity \cite{Bonola1955, Bennett2011,Clifton2006,Sotiriou2010,Capozziello:2025hyw}. On the theoretical side, GR has not been reconciled with quantum field theory in a complete quantum theory of gravity, and its classical dynamics predicts spacetime singularities, indicating the breakdown of the classical description and motivating more fundamental frameworks \cite{Hooft1974,Obukhov_2017,Giulini2003,Rovelli2004}. On the observational side, the standard $\Lambda$CDM model fits a wide range of data but relies on ingredients whose microphysical origin remains unclear \cite{Perivolaropoulos:2021jda,DESI:2024mwx}, such as dark matter \cite{Mayet:2016zxu,Capozziello:2017rvz,Arun_2017}, dark energy \cite{Carroll:2000fy,Planck:2018vyg}, and the mechanism driving inflation \cite{Guth:2005zr}. It also faces persistent precision tensions, such as the $H_0$ and $\sigma_8$ tensions \cite{Aloni:2021eaq,DiValentino:2021izs,Pantos:2026koc}. These tensions have become central data-driven tests of the concordance model and provide a concrete observational impetus for exploring new gravitational dynamics.

The arena we will use for such explorations is the metric-affine framework, where the metric $g_{\mu\nu}$ and the affine connection $\Gamma^{\rho}{}_{\mu\nu}$ are treated as independent geometric structures \cite{Ferraris1982,HEHL19951}. In standard GR, one assumes vanishing torsion and metric compatibility, which uniquely selects the Levi-Civita connection \cite{Einstein1916}. Dropping one or both assumptions allows for torsion and/or nonmetricity as additional geometric fields \cite{BeltranJimenez:2019odq}. This broader viewpoint contains alternative, and sometimes classically equivalent, formulations of GR: teleparallel gravity attributes gravity to torsion with vanishing curvature, while symmetric teleparallel gravity attributes gravity to nonmetricity with vanishing curvature and torsion \cite{Aldrovandi2013,Nester:1998mp}. 

We are motivated to study a gravitational theory built from an action containing quadratic combinations of the nonmetricity tensor, known as Newer General Relativity \cite{BeltranJimenez:2017tkd}. This theory is named in analogy with a similar construction based on quadratic combinations of the torsion tensor called New General Relativity \cite{Hayashi1979,Blixt}. In Newer GR, the vanishing of curvature and torsion determines the connection up to an arbitrary gauge field $\xi^{a}$ representing a change of coordinates, which can be chosen in such a way as to trivialize the connection, leading to the so-called coincident gauge. Several aspects of this theory have been studied, including the propagation of gravitational waves \cite{Hohmann:2018wxu}, the post-Newtonian limit \cite{Flathmann:2020zyj}, spherically symmetric solutions \cite{Hohmann:2024phz}, the classification of primary constraints \cite{Dambrosio:2020wbi}, the weak gravity limit \cite{Golovnev:2024owe}, and cosmology \cite{Hohmann:2025kxp}, among others \cite{Tomonari:2025axd,Ganz:2025ydt,Chen:2025mhu}. Current evidence indicates that there are two one-parameter families of Newer GR models with a post-Newtonian limit compatible with solar-system observations, which may also be free of instabilities and viable as gravitational theories \cite{HohmannPN}. 

However, the number of propagating degrees of freedom of the most general Newer GR models, including their one-parameter families, is not yet fully understood. Determining this number is crucial for justifying and characterizing statements about strong coupling issues, ghosts, and other possible instabilities that are relevant in gravity theories \cite{ErrastiDiez:2023gme,ErrastiDiez:2024hfq}. Similar issues have been identified in modified metric teleparallel theories such as $f(T)$ gravity, where Hamiltonian analysis reveals extra degrees of freedom that do not all manifest at the perturbative level  \cite{Li:2011rn,Ferraro:2018axk,Guzman:2019oth,Ferraro:2020tqk,Blagojevic:2020dyq,Golovnev:2020nln,Golovnev:2020zpv}

In gravity and other gauge theories, it is often difficult to determine the physically relevant content from the field equations alone, due to the interplay between dynamical variables and gauge redundancies. This information is encoded in the constraint structure of the theory. Although recent proposals for determining the constraints and degrees of freedom of first-order field theories are formulated at the Lagrangian level \cite{ErrastiDiez:2020dux,ErrastiDiez:2023gme}, their consistency must preferably be tested through the Dirac-Bergmann algorithm. 
This algorithm provides a systematic diagnostic: primary constraints signal a singular Legendre transform, secondary (and higher) constraints follow from demanding preservation of primary constraints under time evolution, and the classification into first- and second-class constraints determines the number of propagating degrees of freedom as well as the generators of gauge symmetries on phase space. For non-Riemannian gravity theories, this analysis is particularly decisive because an independent connection introduces additional gauge freedoms as well as additional constraint surfaces, whose consistency controls whether extra modes are eliminated or become pathological. This analysis could also clarify the role of projective symmetries in the action from projective transformations in the connection. Other ways of interpreting degrees of freedom in modified gravities consist in changes of variables, such as conformal and disformal transformations \cite{Golovnev:2019kcf}. 

In this work we will start the first steps of the Dirac-Bergmann algorithm for Newer General Relativity, that is, to obtain the canonical momenta, and the conditions of invertibility of velocities by computing the kernel of the Hessian. The presence of arbitrary coefficients in the Lagrangian spans several different possibilities for the possible number of primary constraints, which later in the algorithm give rise to different branches, each one with different primary Hamiltonian, time evolution of constraints, and therefore number of degrees of freedom. To our knowledge, only Ref.\cite{Dambrosio:2020wbi} treats the classification of primary constraints in Newer GR, and Ref. \cite{Bajardi:2024qbi} performs a more general analysis allowing torsion in the geometry in addition to nonmetricity. 

The paper is organized as follows. In Sec. \ref{sec:geometry} we introduce the symmetric teleparallel geometry. In Sec. \ref{sec:canonical} we discuss the different possibilities for defining the canonical variables. In Sec. \ref{sec:spectral} we introduce the spectral decomposition approach on the nonmetricity tensor as a tool for finding primary constraints. In Sec. \ref{sec:irreducible} we analyze the Hessian, compute its kernel, and find all primary constraints using an irreducible decomposition. In Sec. \ref{sec:models} we classify all possible models based on the number of primary constraints. In Sec. \ref{sec:cpc} we propose different approaches for considering the symmetric teleparallel connection into the Hamiltonian analysis, and finally we present our conclusions in Sec. \ref{sec:concl}.

\section{The geometric arena for symmetric teleparallel gravity}
\label{sec:geometry}

Consider a spacetime manifold equipped with two geometric, independent entities, a metric $g_{\mu\nu}$ and an affine connection $\Gamma^{\rho}{}_{\mu\nu}$. In a general metric-affine framework, the connection is not required to be compatible with the metric, nor does it have to be symmetric. The extent to which parallel transport fails to preserve the metric is characterized by the nonmetricity tensor,
\begin{equation}
Q_{\rho\mu\nu} \equiv \nabla_{\rho} g_{\mu\nu} = \partial_{\rho} g_{\mu\nu} - \Gamma^{\alpha}{}_{\mu\rho} g_{\alpha \nu} - \Gamma^{\alpha}{}_{\nu\rho} g_{\mu\alpha}.
\label{nonmetr}
\end{equation}
Meanwhile, the antisymmetric component of the connection gives rise to the torsion tensor,
\begin{equation}
T^{\rho}{}_{\mu\nu} \equiv -2\Gamma^{\rho}{}_{[\mu\nu]} =  \Gamma^{\rho}{}_{\nu\mu} -\Gamma^{\rho}{}_{\mu\nu}.
\end{equation}
The curvature associated with the affine connection is described by the Riemann tensor,
\begin{equation}
{R}{}^{\rho}{}_{\lambda\mu\nu} = \partial_{\mu}\Gamma^{\rho}{}_{\lambda\nu} - \partial_{\nu}\Gamma^{\rho}{}_{\lambda\mu} + \Gamma^{\rho}{}_{\sigma\mu}\Gamma^{\sigma}{}_{\lambda\nu} - \Gamma^{\rho}{}_{\sigma\nu}\Gamma^{\sigma}{}_{\lambda\mu}.
\end{equation}
In this work we restrict ourselves to the torsionless sector 
\begin{equation}
T^{\rho}{}_{\mu\nu}=0 \qquad \Longleftrightarrow \qquad \Gamma^{\rho}{}_{\mu\nu}=\Gamma^{\rho}{}_{\nu\mu},
\end{equation}
so that the independent connection is symmetric. Additionally, we impose that the Riemann tensor of the affine connection vanishes, which determines the connection as
\begin{equation}
    \Gamma^{\rho}{}_{\mu\nu} = \dfrac{\partial x^{\rho}}{\partial \xi^{a} } \dfrac{\partial^2 \xi^{a}}{\partial x^{\mu} \partial x^{\nu}},
    \label{symtelcon}
\end{equation}
where $\xi^{a}(x^{\mu})$ is a collection of four scalars representing a change of coordinates from $x^{\mu}$, with $\frac{\partial \xi^{a}}{\partial x^{\mu}}$ the (invertible) coordinate transformation matrix. 

Since the metric is symmetric, the nonmetricity tensor satisfies $Q_{\rho\mu\nu}=Q_{\rho(\mu\nu)}$, and it admits the two independent traces
\begin{equation}
Q_{\mu}:=Q_{\mu\nu}{}^{\nu}, \qquad \tilde Q_{\mu}:=Q^{\nu}{}_{\mu\nu}.
\end{equation}
At lowest order in derivatives, and restricting to parity-even terms, the most general gravitational action built from all quadratic scalars in the nonmetricity tensor is 
\begin{equation}
\begin{split}
S (g_{\mu\nu}, \xi^{a}) & =   \int d^4x \mathcal{L} = \dfrac12 \int d^4x \sqrt{-g} \mathbb Q  \\ & = \dfrac12 \int d^{4}x \sqrt{-g} \left[  c_1 Q_{\rho\mu\nu} Q^{\rho\mu\nu} + c_2 Q_{\nu\mu\rho} Q^{\rho\mu\nu}  + c_3 Q^{\mu}Q_{\mu} \right. \\
& \left.+ c_4 \tilde{Q}^{\mu} \tilde{Q}_{\mu} + c_5 Q^{\mu} \tilde{Q}_{\mu} \right],
\end{split}
\label{newergraction}
\end{equation}
where the nonmetricity scalar $\mathbb Q$ is implicitly defined, and the coefficients $c_{1},\ldots,c_{5}$ control the five independent quadratic nonmetricity invariants. We fall into the symmetric teleparallel equivalent of general relativity case when $c_1=-1/4$, $c_2 = 1/2$, $c_3= 1/4$, $c_4 = 0$, $c_5 = -1/2 $, which reproduce the usual nonmetricity scalar of STEGR \cite{Nester:1998mp,Adak:2004uh,Adak:2005cd,Adak:2006rx,Adak:2026mgj,Capozziello:2022zzh}.

For the nonmetricity scalar it is commonly used in the literature in the form 
\begin{equation}
\mathbb Q = \mathcal{P}^{\rho\mu\nu}  Q_{\rho\mu\nu} 
\label{Qpq}
\end{equation}
with the non-metricity conjugate or superpotential $\mathcal{P}$ defined as
\begin{equation}
\mathcal{P}^{\rho\mu\nu} =  c_1 Q^{\rho\mu\nu} + c_2 Q^{(\nu\mu)\rho}  + c_3 Q^{\rho} g^{\mu\nu} + c_4   g^{\rho(\nu} \tilde{Q}^{\mu)} + \frac{c_5}{2}\left( g^{\rho(\nu} Q^{\mu)} + \tilde{Q}^{\rho}g^{\mu\nu}\right) .
\end{equation}

It will be convenient for later to alternatively collect the quadratic nonmetricity terms into
\begin{equation}
\mathbb Q = \tilde{P}^{\alpha\beta\gamma\,\rho\mu\nu} Q_{\alpha\beta\gamma } Q_{\rho\mu\nu}
\label{Qpq2}
\end{equation}
where
\begin{equation}
\tilde{P}^{\alpha\beta\gamma\,\rho\mu\nu}  =
c_{1} g^{\alpha\rho}g^{\beta\mu}g^{\gamma\nu}
+c_{2} g^{\alpha\nu}g^{\beta\mu}g^{\gamma\rho}
+c_{3} g^{\alpha\rho}g^{\beta\gamma}g^{\mu\nu}
+c_{4} g^{\alpha\gamma}g^{\beta\mu}g^{\rho\nu}
+c_{5} g^{\alpha\mu}g^{\beta\gamma}g^{\rho\nu}.
\end{equation}
A similar structure is shared with metric teleparallel gravities which has shown beneficial for Hamiltonian analysis \cite{Ferraro:2016wht,Ferraro:2018tpu,Guzman:2020kgh}.

Given the symmetries of the nonmetricity tensor on the last two indices in the previous definition of $\mathbb Q$, it would be convenient to define a fully symmetrized object $P^{\alpha(\beta\gamma)\rho(\mu\nu)} $ such that
\begin{equation}
\begin{split}
P^{\alpha(\beta\gamma)\rho(\mu\nu)} & = c_1 g^{\alpha\rho} g^{\mu(\beta} g^{\gamma)\nu} + c_2 g^{\alpha(\mu} g^{\rho(\gamma} g^{\beta)\nu)} + c_3 g^{\alpha\rho} g^{\beta\gamma} g^{\mu\nu}\\
& + c_4 g^{\alpha(\beta} g^{\gamma)(\mu} g^{\nu)\rho}+ c_5 g^{\beta\gamma} g^{\alpha(\mu} g^{\nu)\rho},
\end{split}
\end{equation}
or explicitly
\begin{equation}
\begin{split}
P^{\alpha(\beta\gamma)\rho(\mu\nu)} & = \dfrac{c_1}{2} g^{\alpha\rho} ( g^{\beta\mu} g^{\gamma\nu} + g^{\beta\nu} g^{\gamma\mu} ) \\
& + \dfrac{c_2}{4}( g^{\alpha\nu}g^{\beta\mu}g^{\gamma\rho} + g^{\alpha\nu} g^{\gamma\mu}g^{\beta\rho} + g^{\alpha\mu} g^{\beta\nu} g^{\gamma\rho} + g^{\alpha\mu} g^{\gamma\nu} g^{\beta\rho} )\\
& + c_3 g^{\alpha\rho} g^{\beta\gamma} g^{\mu\nu} \\
& + \dfrac{c_4}{4} ( g^{\alpha\gamma} g^{\beta\mu} g^{\rho\nu} + g^{\alpha\beta} g^{\gamma\mu} g^{\rho\nu} + g^{\alpha\gamma} g^{\beta\nu} g^{\rho\mu} + g^{\alpha\beta} g^{\gamma\nu} g^{\rho\mu} ) \\
& + \dfrac{c_5}{2} g^{\beta\gamma} ( g^{\alpha\mu} g^{\rho\nu} + g^{\alpha\nu} g^{\rho\mu} ).
\end{split}
\end{equation}
This object then satisfies the following symmetries 
\begin{equation}
P^{\alpha(\beta\gamma) \rho(\mu\nu)} = P^{\alpha(\beta\gamma) \rho(\nu\mu)} = P^{\alpha(\gamma\beta) \rho(\mu\nu)} = P^{\alpha(\gamma\beta)\rho(\nu\mu)}.
\end{equation}
With it, the Lagrangian is rewritten by the replacement $\tilde{P}^{\alpha\beta\gamma\rho\mu\nu} \longrightarrow P^{\alpha(\beta\gamma)\rho(\mu\nu)}$.

The metric equations of motion of Newer GR are obtained from taking the variation of \eqref{newergraction} with respect to the metric, giving 
\begin{equation}
\label{metriceqns}
\dfrac{2}{\sqrt{-g} } \nabla_{\rho} ( \sqrt{-g} \mathcal P^{\rho\mu\nu} ) + \mathcal P^{\mu\rho\lambda} Q^{\nu}{}_{\rho\lambda} + 2 \mathcal P^{\rho\lambda\mu} Q_{\rho\lambda}{}^{\nu} - \mathbb Q g^{\mu\nu} = 0
 \end{equation}
in vacuum. Taking the variation of \eqref{newergraction} with respect to the connection gives \cite{Blixt:2023kyr}
\begin{equation}
\nabla_{\rho} \nabla_{\mu} \left(\sqrt{-g} \mathcal{P}^{\rho \mu}{}_{\nu} \right) = 0,
\label{conecteqn}
\end{equation}
in the absence of hypermomentum.

It is not hard to see that the metric equations \eqref{metriceqns} are 3rd order in derivatives in $\xi^{a}$, while they appear in \eqref{conecteqn} with derivatives of 4th order, while being up to derivatives of 3rd order in the metric tensor.

\section{The canonical variables in symmetric teleparallel gravity}
\label{sec:canonical}

A symmetric teleparallel gravity theory is defined in terms of two sets of fields: the metric $g_{\mu\nu}$ and the four scalar fields $\xi^{a}(x^{\mu})$ determining the affine connection through \eqref{symtelcon}. However, as noted in \cite{Blixt:2023kyr}, the latter expression contains second-order derivatives in the $\xi^{a}$ fields. Therefore, if they are taken as fundamental variables, the theory might appear prone to Ostrogradsky instabilities, and the connection equation becomes fourth-order in the fields $\xi^{a}$. This fact has been overlooked in the literature \cite{BeltranJimenez:2017tkd}, and has been avoided either by imposing the coincident gauge, through the use of Lagrange multipliers imposing zero torsion and curvature, or by writing the equations directly in terms of the connection. 

The canonical momenta are generally obtained by differentiating the Lagrangian with respect to the velocities. In a metric-affine theory, the canonical variables are $g_{\mu\nu}$ and $\Gamma^{\rho}{}_{\mu\nu}$. Therefore, we have
\begin{equation}
    \pi^{\mu\nu} = \dfrac{\partial \mathcal{L}}{ \partial \dot{g}_{\mu\nu}}, \qquad  \Pi_{\rho}{}^{\mu\nu} = \dfrac{\partial \mathcal{L}}{\partial \dot{\Gamma}^{\rho}{}_{\mu\nu}},
    \label{momgG}
\end{equation}
where dots denote time derivatives. Later we will also use the notation $\partial_0$ for time derivatives interchangeably when convenient. Here we temporarily treat the connection as an independent field, although, as mentioned before, this is not the case in symmetric teleparallel geometry. We focus on obtaining the primary constraints arising from the metric momenta, and the inclusion of the symmetric teleparallel connection should not affect the number of primary constraints in the metric sector that we derived below.

From the standard definition of the nonmetricity tensor $Q_{\rho\mu\nu}$ we observe that the dependence on $\dot{g}_{\mu\nu}$ comes only from the $\rho=0$ component, since
\begin{equation}
    Q_{0\mu\nu} = \dot{g}_{\mu\nu} - \Gamma^{\alpha}{}_{\mu 0} g_{\alpha\nu} - \Gamma^{\alpha}{}_{\nu 0} g_{\mu\alpha},
\end{equation}
which implies that
\begin{equation}
   \dfrac{\partial Q_{\rho\mu\nu}}{\partial (\partial_0 {g}_{\alpha \beta}) } = \delta^{0}_{\rho} \delta^{\alpha}_{(\mu} \delta^{\beta}_{\nu)}.
\end{equation}
Using the chain rule $\frac{\partial}{\partial \dot{g}_{\alpha\beta}} = \frac{\partial}{\partial Q_{0\mu\nu}} \frac{\partial Q_{0 \mu\nu}}{ \partial \dot{g}_{\alpha\beta}} $ to compute this variation in the Lagrangian in \eqref{newergraction}, we obtain
\begin{equation}
\pi^{\alpha\beta} = \dfrac{1}{2} \sqrt{-g} \dfrac{\partial \mathbb Q}{\partial Q_{0\mu\nu}} \delta^{\alpha}_{(\mu} \delta^{\beta}_{\nu)} .
\end{equation}
Using the form \eqref{Qpq} of $\mathbb Q$, this variation gives simply
\begin{equation}
\pi^{\mu\nu} = \dfrac12 \sqrt{-g} \mathcal P^{0\mu\nu}.
\label{momnew}
\end{equation}
If, instead, one wants the explicit dependence of the nonmetricity tensor on the canonical momenta, we must consider the expression \eqref{Qpq2} for the nonmetricity scalar. Taking into account the symmetrization of the tensor $P^{\alpha\beta\gamma\rho\mu\nu}$, we write the canonical momenta as \footnote{The symmetrization works as $A_{(ij)} = \frac12(A_{ij} + A_{ji} )$.}
\begin{equation}
\pi^{\alpha\beta} = \dfrac{\sqrt{-g}}{2} \left( P^{0(\alpha\beta)\rho(\mu\nu)} + P^{\rho(\mu\nu) 0 (\alpha\beta)} \right) Q_{\rho\mu\nu}.
\end{equation}
We expand this expression to find the dependence of the momenta on the velocities
\begin{equation}
\label{mom-vel}
\pi^{\alpha\beta} = \dfrac{1}{2} \sqrt{-g}\left( K^{\alpha\beta\mu\nu} Q_{0 \mu\nu} + B^{\alpha\beta} \right)
\end{equation}
where
\begin{equation}
\label{HessP}
K^{\alpha\beta\mu\nu} = P^{0(\alpha\beta)0(\mu\nu)} + P^{0(\mu\nu) 0 (\alpha\beta)}
\end{equation}
is the Hessian of the Lagrangian with respect to the velocities, and
\begin{equation}
B^{\alpha\beta} = \left( P^{0(\alpha\beta)i(\mu\nu)} + P^{i(\mu\nu) 0 (\alpha\beta)} \right) Q_{i\mu\nu}
\label{Bmostgeneral}
\end{equation}
encode terms independent of the velocities.
In \eqref{mom-vel}, we can go further and extract the components of the connection in the nonmetricity tensor that do not contain time derivatives. This means that we can rewrite \eqref{nonmetr} as
\begin{equation}
Q_{\rho\mu\nu} = \partial_{\rho} g_{\mu\nu} - q_{\rho\mu\nu}
\end{equation}
with $q_{\rho\mu\nu} = \Gamma^{\alpha}{}_{\mu\rho} g_{\alpha\nu} + \Gamma^{\alpha}{}_{\nu\rho} g_{\mu\alpha} $, so the momenta become
\begin{equation}
    \pi^{\alpha\beta} = \dfrac{1}{2} \sqrt{-g}\left( K^{\alpha\beta\mu\nu} \partial_0 g_{\mu\nu} + C^{\alpha\beta} \right)
\label{eq:momfull}
\end{equation}
with
\begin{equation}
    C^{\alpha\beta} = B^{\alpha\beta} - K^{\alpha\beta\mu\nu} q_{0\mu\nu}.
\label{eq:momspatial} 
\end{equation} 
In order to compare our results more easily with previous works \cite{Dambrosio:2020wbi,Bajardi:2024qbi}, we introduce the ADM decomposition of the  metric, and later define the Hessian in a different way, in terms of the time derivatives of the ADM variables. 

We consider the ADM decomposition of the metric as \cite{alcubierre2008introduction}
\begin{equation}
g_{00} = - \alpha^2 + \beta_i \beta^i, \qquad g_{0i} = \beta_i, \qquad g_{ij} = \gamma_{ij},
\label{admdown}
\end{equation}
and the inverse metric
\begin{equation}
g^{00}=-\frac{1}{\alpha^2},\quad g^{0i}=\frac{\beta^i}{\alpha^2},\quad g^{ij}=\gamma^{ij}-\frac{\beta^i\beta^j}{\alpha^2}.
\label{admup}
\end{equation}
It also follows that $\sqrt{-g} = \alpha \sqrt{\gamma}$, with $g$ and $\gamma$ the determinants of $g_{\mu\nu}$ and $\gamma_{ij}$. In the coordinate system introduced above, the normal vector has components
\begin{equation}
    n_{\mu} = (-\alpha,0), \qquad n^{\mu} = (1/\alpha, -\beta^i/\alpha).
\end{equation}
With the previous definitions, notice that we can write $g^{0\mu} = -n^{\mu}/\alpha$. It is also relevant to show how the time derivatives of the two sets of variables are related:
\begin{equation}
\label{veloctoadm}
\dot{g}_{00} = -2 \alpha \dot{\alpha} + 2 \beta^{i} \dot{\beta}_{i} - \beta^{i} \beta^{j} \dot{\gamma}_{ij}, \qquad \dot{g}_{0i} = \dot{\beta}_{i}, \qquad \dot{g}_{ij} = \dot{\gamma}_{ij}.
\end{equation}
The ADM variables lead to canonical momenta that differ slightly from those introduced in \eqref{momgG}. Firstly, we denote the canonical momenta conjugated to the ADM variables $\alpha, \beta_{i}$ and $\gamma_{ij}$ as $\overset{\alpha}{\pi},\overset{\beta}{\pi}{}^{i} $ and $ \overset{\gamma}{\pi}{}^{ij}$, respectively. Since the metric components and their inverse depend on the ADM variables as written in \eqref{admdown} and \eqref{admup}, this creates a nontrivial dependence of the ADM momenta on the $\pi^{\mu\nu}$ that can be computed through the following formulas 
\begin{equation}
\label{lapse00}
\overset{\alpha}{\pi} = \dfrac{\delta \mathcal{L}}{\delta \dot{\alpha}} = \dfrac{\delta \mathcal{L}}{\delta \dot{g}_{00} } \dfrac{\delta \dot{g}_{00}}{\delta \dot{\alpha}} = - 2\alpha \dfrac{\delta \mathcal{L}}{\delta \dot{g}_{00}} = -2\alpha \pi^{00},
\end{equation}
\begin{equation}
\label{shift0i}
  \overset{\beta}{\pi}{}^{i} =  \dfrac{\delta \mathcal{L}}{\delta \dot{\beta}_i} = \dfrac{\delta \mathcal{L}}{\delta \dot{g}_{00} } \dfrac{\delta \dot{g}_{00}}{\delta \dot{\beta}_i} + \dfrac{\delta \mathcal{L}}{\delta \dot{g}_{0j} } \dfrac{\delta \dot{g}_{0j}}{\delta \dot{\beta}_i} = 2 (\gamma^{ij} \beta_j \pi^{00} + \pi^{0i}),
\end{equation}
\begin{equation}
\label{spmomrel}
  \overset{\gamma}{\pi}{}^{ij} = \dfrac{\delta \mathcal{L}}{\delta \dot{\gamma}_{ij} } = \dfrac{\delta \mathcal{L}}{\delta \dot{g}_{00} } \dfrac{\delta \dot{g}_{00}}{\delta \dot{\gamma}_{ij}} + \dfrac{\delta \mathcal{L}}{\delta \dot{g}_{kl} } \dfrac{\delta \dot{g}_{kl}}{\delta \dot{\gamma}_{ij}} = - \beta^{i} \beta^{j} \pi^{00} + \pi^{ij}.
\end{equation}
The transformations of momenta given above do not depend on a particular gravitational Lagrangian. In the most general cases where there is no gauge invariance, it is the only form these relations could have. However, special cases, as for instance STEGR, might behave differently. In this specific case, the $\pi^{00}$ momentum is constrained but does not vanish, in contrast to the situation in the Dirac or ADM formulations of GR. Had we started from the ADM formulation of STEGR directly, we would have obtained an explicit expression for the temporal momentum in the relation \eqref{spmomrel} rather than $\pi^{00}$ itself, leading to a different symplectic structure in the phase space.

Such subtleties have led some authors \cite{Kiriushcheva:2010ycc,Frolov:2022qvz} to claim that the ADM formalism is not a proper formulation of GR, since the associated change of variables is not canonical. There is a valid point behind this argument: as long as one insists on working in Darboux variables, the transformation must be canonical \cite{Darboux}. The corresponding change of the momenta \cite{Golovnev:2022rui} is given above. When applied to the Dirac formulation, it is only equivalent on-shell to what one would have obtained by starting directly from the ADM parametrization of GR. The latter approach effectively corresponds to the transformation without explicit $\pi^{00}$ and $\pi^{0i}$ terms which is indeed not canonical. This does not mean that the standard ADM formulation of GR is wrong, but simply reflects the fact that gauge theories do not have a unique Hamiltonian formulation, with reparameterization of Lagrange multipliers available and welcome.

This non-uniqueness is true for any constrained system. The definitions of the momenta might be shifted by combinations of primary constraints. This corresponds to the freedom to modify the canonical Hamiltonian by adding combinations of those primary constraints to it, effectively reparametrizing the Lagrange multipliers in the total Hamiltonian. In the second-class case, this is compensated by the corresponding change in the equations that fix the multipliers through preservation of the secondary constraints. In the first-class case, this freedom is the essence of gauge symmetry \cite{Prokhorov:2011zz}.

\section{Classification of Newer GR models in a spectral approach}
\label{sec:spectral}

In this section, our aim is to classify models in Newer GR according to the presence of primary constraints. It has been established in the literature that if different combinations of the parameters $c_i$ vanish, then the number of primary constraints changes depending on the combinations, which leads to qualitatively different models with different physics.

For analyzing the constraints, we consider the momenta defined in \eqref{mom-vel},
\begin{equation*}
\pi^{\alpha\beta} = \dfrac{1}{2} \sqrt{-g}\left( K^{\alpha\beta\mu\nu} Q_{0 \mu\nu} + B^{\alpha\beta} \right),
\end{equation*}
with particular attention to the first term in this expression. The relevant problem is to determine the structure of the kernel of the linear operator $\dfrac{1}{2} \sqrt{-g} K^{\alpha\beta\mu\nu}$ that maps the space of  variables $Q_{0\mu\nu}$ to the space of momenta $\pi^{\mu\nu}$.

This operator is represented by a symmetric rank-4 tensor $K^{\alpha\beta\mu\nu}=K^{\mu\nu\alpha\beta}$, but it is possible to map it to a symmetric matrix once the pairs of symmetric indices are combined into a unique index of ten values in the usual spacetime dimension. In other words, we can order a tuple of indices $\mu\nu$ into a list labeled by a unique index $A$ \footnote{A similar treatment for a tensorial Hessian has been done for TEGR in \cite{Ferraro:2016wht}.}. Since the resulting matrix is real and symmetric, it can always be diagonalized. However, this direct diagonalization is not the most convenient procedure in the present setting, because the domain and codomain carry indices in different positions. It is useful for identifying the kernel, but less natural for obtaining the full decomposition of the operator.

Our approach will be in slightly changing the shape of the momenta-velocities relation to
\begin{equation}
\label{relQpi}
\pi^{\alpha\beta} = \dfrac{1}{2} \sqrt{-g}\left( K^{\alpha\beta\mu\nu} g_{\mu\rho} g_{\nu\sigma} Q_{0}^{\hphantom{0} \rho\sigma} + B^{\alpha\beta} \right),
\end{equation}
without modifying the tensor $B^{\alpha\beta}$ that denotes the terms depending on $Q_{i\mu\nu}$ and does not contain any metric velocities. In this form, we will see that the full spectral decomposition of the operator in the first term is easily available. 

\subsection{Linear algebra considerations}

Once we have changed the variables encoding the velocities to $Q_{0}^{\hphantom{0} \mu\nu}$, the matrix at hand \eqref{relQpi} given by $\dfrac{1}{2} \sqrt{-g} K^{\alpha\beta\mu\nu} g_{\mu\rho} g_{\nu\sigma}$, is no longer symmetric. Let us show, however, that its corresponding linear operator is still symmetric, in a well-defined meaning of this word. Moreover, since we are in finite dimensions, it is actually a self-adjoint operator.

For simplicity, consider a dynamical theory containing a 1-form field $v_{\alpha}$ and a Lorentzian metric $g_{\mu\nu}$. Let the kinetic part of its Lagrangian be $\frac12 H^{\alpha\beta} {\dot v}_{\alpha}{\dot v}_{\beta}$, so that the canonical momenta are 
\begin{equation}
\pi^{\alpha} =H^{(\alpha\beta)}{\dot v}_{\beta}
\end{equation}
and the matrix $H$ should naturally be taken symmetric from the very beginning. However, naively studying the spectrum of the operator $H$ would make us use relations like $\zeta^{\alpha} = \lambda \zeta_{\alpha}$ for an eigenvector $\zeta$ with an eigenvalue $\lambda$. It is then more convenient to analyze the operator $H$ in the form
\begin{equation}
\pi^{\alpha}=H^{\alpha\beta}g_{\beta\gamma}{\dot v}^{\gamma} \end{equation}
with ${\dot v}^{\gamma} \equiv g^{\gamma\beta} {\dot v}_{\beta}$ analogous to our velocity variable $Q_{0}^{\hphantom{0} \mu\nu}$,
and study the spectrum of an asymmetric matrix $H^{\alpha\beta}g_{\beta\gamma}$.

At the same time, if we consider the domain and the codomain as a vector space with an inner product defined by the metric $g$, and denote the operator by $\mathfrak H$, then one can easily find that
\begin{equation}
{\overrightarrow \pi}\cdot{\mathfrak H} \overrightarrow{\dot v} = \pi^{\alpha} g_{\alpha\beta}\cdot H^{\beta\mu} g_{\mu\nu} {\dot v}^{\nu} = {\dot v}^{\nu} g_{\nu\mu} \cdot H^{\mu\beta}g_{\beta\alpha}\pi^{\alpha} = {\mathfrak H} {\overrightarrow \pi}\cdot \overrightarrow{\dot v},
\end{equation}
and therefore the operator is symmetric. The only difference from the usual linear algebra cases is that it acts in an indefinite inner-product space.

Unfortunately, the spectral analysis in indefinite inner-product spaces is much more complicated in many respects, see for example Ref. \cite{bognar1974indefinite}. However, we will show that it is still possible to have complete information on the spectral properties of the operator $K^{\alpha\beta\mu\nu} g_{\mu\rho} g_{\nu\sigma}$. Note also that the constraints come with some eigenvalues being zero, as the image would not then cover the full momentum space. One can think of that as a cokernel of the operator.

\subsection{Previous results and problems}

The canonical momenta can be obtained in a different form than  \eqref{mom-vel} by direct variation of the Lagrangian as
\begin{multline}
\label{momindup}
\frac{\pi^{\mu\nu}}{\sqrt{-g}} = c_1 Q^{0\mu\nu} + \frac{c_2}{2}\left(Q^{\mu\nu 0}+Q^{\nu\mu 0}\right) + c_3 g^{\mu\nu}Q^0\\
+ \frac{c_4}{2}\left(g^{0\mu}{\tilde Q}^{\nu} + g^{0\nu}{\tilde Q}^{\mu}\right) + \frac{c_5}{4} \left(2g^{\mu\nu} {\tilde Q}^{0}+g^{0\mu}Q^{\nu}+g^{0\nu}Q^{\mu}\right)
\end{multline}
and can be rewritten as
\begin{multline}
\label{momindus}
\frac{\pi^{\mu\nu}}{\sqrt{-g}} = c_1 g^{00} Q_0^{\hphantom{0}\mu\nu} + \frac{c_2 + c_4}{2}\left(g^{\mu 0}Q_0^{\hphantom{0}\nu 0}+g^{\nu 0}Q_0^{\hphantom{0}\mu 0}\right) \\ 
+ c_3 g^{\mu\nu} g^{00} g_{\alpha\beta} Q_0^{\hphantom{0}\alpha\beta}
 + \frac{c_5}{2} \left(g^{\mu\nu} Q_0^{\hphantom{0}00}+g^{0\mu}g^{0\nu}g_{\alpha\beta}Q_0^{\hphantom{0}\alpha\beta}\right) + \frac12 B^{\mu\nu}
\end{multline}
where, precisely as above, $B^{\mu\nu}$ stands for terms without velocities. 

Note that, in the kinetic part, the coefficients $c_2$ and $c_4$ appear only through the combination $c_2+c_4$. In the weak gravity limit, the same dependence on $c_2$ and $c_4$ through their sum was found in the full quadratic action \cite{Golovnev:2024owe}. Beyond that limit, this is not the case inside $B^{\mu\nu}$. However, the presence of primary constraints depends on only four independent parameters: $c_1$, $c_2+c_4$, $c_3$, and $c_5$.

We write the various components explicitly:
\begin{equation}
\label{momtemp}
\frac{\pi^{00}}{\sqrt{-g}} = \left(c_1 + c_2 + c_4 + \frac{c_5}{2}\right) g^{00} Q_0^{\hphantom{0}00} + \left(c_3 + \frac{c_5}{2}\right) (g^{00})^2 g_{\alpha\beta} Q_0^{\hphantom{0}\alpha\beta} + \frac12 B^{00},
\end{equation}

\begin{multline}
\label{mommix}
\frac{\pi^{0i}}{\sqrt{-g}} = \left(c_1 + \frac{c_2+c_4}{2}\right) g^{00}Q_0^{\hphantom{0} 0 i}  + \frac{(c_2 + c_4 + c_5)}{2} g^{0i} Q_0^{\hphantom{0}00}\\
+ \left(c_3 +\frac{c_5}{2}\right)g^{0i}g^{00}g_{\alpha\beta} Q_0^{\hphantom{0}\alpha\beta} + \frac12 B^{0i},
\end{multline}

\begin{multline}
\label{momspat}
\frac{\pi^{ij}}{\sqrt{-g}}  = c_1 g^{00} Q_0^{\hphantom{0}ij} + \frac{c_2 + c_4}{2} \left(g^{0i}Q_0^{\hphantom{0}0j} + g^{0j}Q_0^{\hphantom{0}0i}\right) +\frac{c_5}{2} g^{ij} Q_0^{\hphantom{0}00}\\
 + \left(c_3 g^{ij}g^{00}+\frac{c_5}{2} g^{0i}g^{0j}\right) g_{\alpha\beta}Q_0^{\hphantom{0}\alpha\beta} + \frac12 B^{ij}.
\end{multline}

The cases in which different numbers of primary constraints appear for this system were analyzed in \cite{Dambrosio:2020wbi}. In Refs. \cite{Dambrosio:2020wbi,DAmbrosio:2020nqu}, the authors emphasize as a novel feature of the STEGR action that the momenta conjugate to the lapse and shift are nonvanishing, although constrained. However, this feature was already known to Dirac in the form of nonvanishing components $\pi^{0i}$. Indeed, the STEGR action is equivalent to the $\Gamma\Gamma$ action of GR, and Dirac already observed these non-zero momenta in his 1958 paper \cite{Dirac:1958sc}. This is precisely one of the motivations for adding a boundary term to the action, so as to make these momenta vanish. Nevertheless, Dirac explicitly noted that the corresponding momenta are constrained even before this modification.

We should also point out several inconsistencies in \cite{Dambrosio:2020wbi}. For example, in Section 3 the authors correctly state that, in STEGR, the momentum conjugate to the lapse is constrained. However, the temporal component of their Eq. (3.4), or Eq. (14) in the arXiv version, implies the opposite. The reason for this discrepancy can be traced back to their Eq. (3.2) (or  Eq. (12) in arXiv version), where the same component already contains an error. The lapse momentum, given in our notation as \eqref{lapse00}, is proportional to $\pi^{00}$. The coefficients multiplying the nonmetricity traces agree with those in our expression \eqref{momindup}. However, apart from the contributions contained in the traces themselves, the coefficient in front of $Q^{000}$ should be $c_1+c_2$, whereas in \cite{Dambrosio:2020wbi} it is written as $c_1+c_2+c_3+c_4+c_5$. 

This error appears to be confined to Section 3. In Section 4, the authors give the condition for their so-called Case I, which coincides with the correct condition for the vanishing of the lapse momentum. Their Cases II and III are also correct, although we will show that they admit a more transparent interpretation. Their Case IV is, however, more problematic. It is obtained from a complicated determinant whose coefficients depend on some metric components. In order for this determinant to vanish throughout the entire phase space, the authors set independently to zero the coefficients multiplying two different quantities. If their calculation were otherwise correct, this would suggest that their analysis may miss possible strong coupling issues already at the level of the primary constraints.

We will show that both Cases I and IV in Ref. \cite{Dambrosio:2020wbi} are particular cases of a more general quadratic equation which sets a determinant of a $2\times 2$ matrix to zero. In particular, the case of a constraint for
\begin{equation}
\label{newconst}
\frac{g_{\mu\nu}\pi^{\mu\nu}}{\sqrt{-g}} = \left(c_1 + 4c_3 +\frac{c_5}{2}\right)g^{00} g_{\alpha\beta} Q_0^{\hphantom{0}\alpha\beta} + \left(c_2 + c_4 + 2c_5\right) Q_0^{\hphantom{0}00} + \frac12 g_{\mu\nu} B^{\mu\nu}
\end{equation}
was missed in \cite{Dambrosio:2020wbi}. The appearance of constraints does not depend on the formalism, that is, whether they are computed in terms of $\pi^{\mu\nu}$ or with the momenta in the ADM formulation. Also, the solvability of the equations for $Q_0^{\hphantom{0}\mu\nu} =g^{\mu\alpha}g^{\nu\beta} Q_{0\alpha\beta}$ is equivalent to the solvability for $Q_{0\mu\nu}$, as long as the metric is nondegenerate, which is always the case.

These expressions can be presented in terms of the metric derivatives, since in the coincident gauge, it is satisfied that
\begin{equation}
{\dot g}^{\mu\nu}=-g^{\mu\alpha}g^{\nu\beta}{\dot g}_{\alpha\beta}=-Q_0^{\hphantom{0}\mu\nu}.
\end{equation}
Therefore, for example, $Q_0^{\hphantom{0}00}=-2\frac{\dot\alpha}{\alpha^3}$ in the ADM variables. We can also easily find the trace of the nonmetricity as $g_{\alpha\beta} Q_0^{\hphantom{0}\alpha\beta} = g^{\alpha\beta} Q_{0\alpha\beta} = 2\frac{\dot\alpha}{\alpha} + \gamma^{ij} {\dot\gamma}_{ij}$.

\subsection{The spectral decomposition}

The explicit relations \eqref{momtemp}, \eqref{mommix}, and  \eqref{momspat} allow us to find three invariant subspaces for the operator $\mathfrak K$ in \eqref{momindus},
\begin{equation}
\frac{\pi^{\mu\nu}}{\sqrt{-g}}={\mathfrak K}^{\mu\nu}_{\alpha\beta} Q_0^{\alpha\beta} + \frac12 B^{\mu\nu}.
\end{equation}
It is possible to unambiguously decompose any rank-2 tensor $Q_0^{\hphantom{0}\mu\nu}$ as
\begin{equation}
\label{newbasis}
Q_0^{\hphantom{0}\mu\nu} = V_{(1)}^{\mu\nu} + V_{(2)}^{\mu\nu} + V_{(3)}^{\mu\nu},
\end{equation}
where $V^{\mu\nu}_{(i)}\in {\mathfrak Q}_{(i)}$, with the subspaces ${\mathfrak Q}_{(i)}$ defined as
\begin{equation}
{\mathfrak Q}_{(1)}: \qquad Q_0^{\hphantom{0}00}=0, \qquad Q_0^{\hphantom{0}0i}=V^i, \qquad Q_0^{\hphantom{0}ij}=\frac{1}{g^{00}}\left(g^{0i}V^j + g^{0j}V^i\right), \
\end{equation}
\begin{equation}
{\mathfrak Q}_{(2)}: \qquad Q_0^{\hphantom{0}00}=Q_0^{\hphantom{0}0i}=0, \qquad Q_0^{\hphantom{0}ij}=M^{ij}, \qquad \gamma_{ij}M^{ij}\equiv 0, \qquad \qquad
\end{equation}
\begin{equation}
{\mathfrak Q}_{(3)}: \qquad Q_0^{\hphantom{0}00}= g^{00}
\Phi, \qquad Q_0^{\hphantom{0}0i}= g^{0i}
\Phi, \qquad Q_0^{\hphantom{0}ij}= \frac{g^{0i}g^{0j}}{g^{00}}
\Phi +\Psi\gamma^{ij}.
\end{equation}
For further use, we denote $V_{(3)}^{\mu\nu}$ as a column vector $\begin{pmatrix} \Phi \\ \Psi \end{pmatrix}$ with the two scalars appearing in the decomposition above. The tensor $\gamma^{ij}$ is the inverse of the spatial metric $\gamma_{ij}=g_{ij}$ as in \eqref{admup}, and one can easily check that $g^{ij}=\gamma^{ij}+ \frac{g^{0i} g^{0j}}{g^{00}}$.

After projecting the momenta onto the spectral basis introduced above, the action of the Hessian takes the form
\begin{multline}
\label{spec}
\frac{{\mathfrak K}^{\mu\nu}_{\alpha\beta} Q_0^{\alpha\beta}}{g^{00}}=\left(c_1 + \frac{c_2 + c_4}{2}\right) V_{(1)}^{\mu\nu} + c_1 V_{(2)}^{\mu\nu} \\
+\begin{pmatrix} c_1 + c_2 + c_3 + c_4 + c_5  & \hphantom{000} & 3\left(c_3 + \frac{c_5}{2}\right) \\ {} & {} & {} \\ c_3 + \frac{c_5}{2}  & \hphantom{000} & c_1 + 3c_3 \end{pmatrix}
\begin{pmatrix} \Phi \\ {} \\ \Psi \end{pmatrix},
\end{multline}
where in the last term we write the subspace ${\mathfrak Q}_{(3)}$ in terms of the column vector $(\Phi,\Psi)^T$. Note that only the $2\times 2$ matrix depends on the dimensionality of the spacetime. For a spacetime of dimension $1+d$, one has $g_{\alpha\beta}Q_0^{\hphantom{0}\alpha\beta}=\Phi + d\Psi$, while we have put $d=3$. Otherwise, the decomposition is general and valid for any spacetime dimension.

It is useful to make explicit the role of the trace $Q_0=g_{\alpha\beta}Q_0^{\hphantom{0}\alpha\beta}$. 
The trace vanishes identically in the subspace ${\mathfrak Q}_{(2)}$, and also in the subspace ${\mathfrak Q}_{(1)}$ as follows from the identity $g_{ij}g^{0i}=-g^{00}g_{0j}$. In the subspace ${\mathfrak Q}_{(3)}$, for a spacetime of $(1+3)$ dimensions, one finds
\begin{equation}
g_{00}Q_0^{\hphantom{0}00} + 2g_{0i}Q_0^{\hphantom{0}0i} +g_{ij}Q_0^{\hphantom{0}ij} =\Phi \left(g_{00}g^{00} + (2-1) \left(1-g_{00}g^{00}\right)\right) + 3\Psi = \Phi + 3\Psi.
\end{equation}
The scalar variables $\Phi$ and $\Psi$ do define their independent nontrivial sector in the spectral decomposition.

We have seen in \eqref{spec} that the Hessian operator is diagonalizable almost independently of the coefficients $c_i$. The subspace ${\mathfrak Q}_{(1)}$ is a three-dimensional eigenspace with eigenvalue $\lambda_1= g^{00} \left(c_1 + \frac{c_2 + c_4}{2}\right)$, while the subspace ${\mathfrak Q}_{(2)}$ is a five-dimensional eigenspace with eigenvalue $\lambda_2= g^{00} c_1$. The only nontrivial mixing occurs in the two-dimensional subspace ${\mathfrak Q}_{(3)}$, where the Hessian is represented by a $2\times 2$ matrix in any spacetime dimension. As expected, and as we will show in the next section, the subspaces ${\mathfrak Q}_{(1)}$, ${\mathfrak Q}_{(2)}$ and ${\mathfrak Q}_{(3)}$ will be identified with vector, tensor and scalar sectors. From now on, we will refer indistinctly to them in this notation or with the notation ${\mathfrak Q}_{(i)}$.

The degeneracy conditions associated with ${\mathfrak Q}_{(1)}$ and ${\mathfrak Q}_{(2)}$ are therefore immediately obtained by setting the corresponding eigenvalues to zero. In this language, Cases II and III of Ref. \cite{Dambrosio:2020wbi} correspond to the vanishing of the coefficients multiplying ${\mathfrak Q}_{(1)}$ and ${\mathfrak Q}_{(2)}$, respectively. The remaining scalar degeneracy is determined by the vanishing of the determinant of the scalar block \eqref{spec}:
\begin{equation}
\label{quadreq}
c_1^2 + c_1 \left(c_2 + c_4 + 4c_3 + c_5 \right) +3c_3 \left( c_2 + c_4 \right) - \frac34 c_5^2 =0.
\end{equation}
This equation defines the general locus on which the scalar sector develops a primary constraint, or two primary constraints if the matrix vanishes as a whole.

Several previously identified cases are contained as special subfamilies of \eqref{quadreq}. For example, the sectors referred to as Cases I and IV \cite{Dambrosio:2020wbi} are inside this equation. They are the $\pi^{00}$ momentum constrained and the equation \eqref{quadreq} with $c_1=0$ respectively. However, \eqref{quadreq} is more general than these particular examples. It also contains further scalar-degenerate subfamilies, such as $2c_1 = 3c_5 = -6c_3$, and  the case of constrained $g_{\mu\nu}\pi^{\mu\nu}$ in \eqref{newconst}.
We also note that an equivalent scalar degeneracy condition as \eqref{quadreq} was obtained in the weak gravity limit in Ref. \cite{Golovnev:2024owe}, up to differences in parameter conventions.

Let us finally comment on the relation with the Hessian determinant given in Ref. \cite{Dambrosio:2020wbi}. In that work the determinant is expressed in terms of two field-dependent quantities, denoted there by $A$ and $B$. As written, those expressions are not invariant under relabelings of spatial coordinates, and therefore they cannot be correct. However, if they were related to each other as $B=\frac34 A$, then the determinant condition of Ref. \cite{Dambrosio:2020wbi} would become equivalent to our results including the scalar degeneracy condition \eqref{quadreq}.

\subsection{The primary constraints}

Having characterized the image of the Hessian in the momentum space, the primary constraints can be found from its cokernel. In other words, whenever one of the irreducible blocks of the Hessian becomes degenerate, the corresponding missing directions in momentum space give rise to primary constraints. We discuss the vector, tensor and scalar sectors separately.

Let us first consider the vector degeneracy $c_1+\frac{c_2 + c_4}{2}=0$. In this case, the image of the Hessian can be represented as
\begin{equation}
\pi^{00}= a g^{00}\Phi, \qquad \pi^{0i}= a g^{0i} \Phi,\qquad \pi^{ij} =a \frac{g^{0i} g^{0j}}{g^{00}} \Phi + a \Psi \gamma^{ij} + b M^{ij}; \qquad \gamma_{ij} M^{ij}\equiv 0
\end{equation}
where $a$ and $b$ denote the corresponding coefficients in the scalar and tensor sectors. We see that the spatial part of the codomain is fully covered, with no restriction, while the mixed components are fixed by an otherwise arbitrary temporal component. Hence, the missing vector directions are consequently selected by the combinations
\begin{equation}
g^{00}\pi^{0i}-g^{0i}\pi^{00},
\end{equation}
and the primary constraints can be given explicitly as
\begin{equation}
\label{expconst}
g^{00}\pi^{0i}-g^{0i}\pi^{00} = \frac12 \sqrt{-g} \left(g^{00}B^{0i}-g^{0i}B^{00}\right).
\end{equation}
Using the relation \eqref{shift0i}, this can be equivalently written as a constraint on the momenta conjugate to the shift,
\begin{equation}
\overset{\beta}{\pi}{}^{i} =  2 (\gamma^{ij} \beta_j \pi^{00} + \pi^{0i}) =\frac{2}{g^{00}}\left(-g^{0i} \pi^{00} + g^{00} \pi^{0i}\right).
\end{equation}
Thus the vector degeneracy yields three primary constraints.

We next consider the tensor degeneracy in the case of $c_1=0$, which is more cumbersome. The image of the Hessian is then spanned by configurations of the form
\begin{equation}
\begin{split}
& \pi^{00}= a g^{00}\Phi, \qquad \pi^{0i}= a g^{0i} \Phi + b V^i,\\
& \pi^{ij} =a \frac{g^{0i} g^{0j}}{g^{00}} \Phi + a \Psi \gamma^{ij} + b \frac{1}{g^{00}}\left(g^{0i}V^j + g^{0j} V^i\right),
\end{split}
\end{equation}
which leads to a constraint on
\begin{equation}
\label{vectcomb}
\Pi^{ij}-\frac13 \gamma^{ij} \gamma_{kl} \Pi^{kl}; \qquad \Pi^{ij}\equiv (g^{00})^2 \pi^{ij} - g^{00}\left(g^{0i}\pi^{0j}+g^{0j}\pi^{0i}\right) + g^{0i}g^{0j}\pi^{00}.
\end{equation}

The constraint can be written explicitly in a full analogy to what was computed in the previous case in Eq. \eqref{expconst}. Namely, we define
$$\mathfrak B^{ij} \equiv (g^{00})^2 B^{ij} - g^{00}\left(g^{0i}B^{0j}+g^{0j}B^{0i}\right) + g^{0i}g^{0j}B^{00}, $$
and if $c_1=0$, then there is a constraint written as 
\begin{equation}
\Pi^{ij} - \dfrac13 \gamma^{ij} \gamma_{kl} \Pi^{kl} =\frac12 \sqrt{-g} \left( {\mathfrak B}^{ij} - \dfrac13 \gamma^{ij} \gamma_{kl} {\mathfrak B}^{kl}\right).
\end{equation}
Note that, since the momenta \eqref{momindup} are functions of the nonmetricity tensor components, one can equivalently use the $C^{\mu\nu}$ expressions \eqref{eq:momspatial} instead of the $B^{\mu\nu}$ in the primary constraints, for when we eliminate the $\partial_0 g_{\mu\nu}$ in a combination of momenta, all the $q_{0\mu\nu}$ do also go away being a part of the same $Q_{0\mu\nu}$.

Finally, if to fully remove the remaining 2D space, the image is 
\begin{equation}
\begin{split}
& \pi^{00}= 0, \qquad \pi^{0i}= b V^i, \\
& \pi^{ij} = a M^{ij} + b \frac{1}{g^{00}}\left(g^{0i}V^j + g^{0j} V^i\right); \qquad \gamma_{ij} M^{ij}\equiv 0.
\end{split}
\end{equation}
Any constraints coming from the scalar sector must be constructed from
\begin{equation}
\pi^{00}\qquad \mathrm{and}\qquad \gamma_{ij} \left(\pi^{ij} - \frac{1}{g^{00}}\left( g^{0i}\pi^{0j}+g^{0j}\pi^{0i}\right) \right) = \gamma_{ij} \pi^{ij} + 2 g_{0i} \pi^{0i}.
\end{equation}
In particular, the constraints for $\pi^{00}$ (the Case I of \cite{Dambrosio:2020wbi}) and for $g_{\mu\nu}\pi^{\mu\nu}$ belong to this sector. And all the other cases here can be thought of as constraining combinations of the expressions $\pi^{00}$ and $g_{\mu\nu}\pi^{\mu\nu}$.

In fact, this is an additional way of obtaining the quadratic equation \eqref{quadreq}, which does not need the spectral decomposition. As we have already seen above in Eqs. \eqref{momtemp} and \eqref{newconst}, these two combinations can be written down as
$$\begin{pmatrix} \frac{\pi^{00}}{\sqrt{-g} g^{00}} \\ {} \\ \frac{g_{\mu\nu}\pi^{\mu\nu}}{\sqrt{-g}} \end{pmatrix} = 
\begin{pmatrix} c_1 + c_2 + c_4 + \frac{c_5}{2}  & \hphantom{000} & c_3 + \frac{c_5}{2} \\ {} & {} & {} \\ c_2 + c_4 + 2 c_5  & \hphantom{000} & c_1 + 4c_3 + \frac{c_5}{2}\end{pmatrix}
\begin{pmatrix} Q_0^{\hphantom{0}00} \\ {} \\ g^{00}g_{\alpha\beta}Q_0^{\hphantom{0}\alpha\beta} \end{pmatrix}
+
\begin{pmatrix} \frac{B^{00}}{2g^{00}} \\ {} \\ \frac{g_{\mu\nu}B^{\mu\nu}}{2} \end{pmatrix}.
$$
 The matrix in this relation has the same determinant as above \eqref{quadreq}, 
 $$c_1^2 + c_1 \left(c_2 + c_4 + 4c_3 + c_5 \right) +3c_3 \left( c_2 + c_4 \right) - \frac34 c_5^2.$$ 
 In other words, the corresponding constraints are given by the cokernel of this simple $2 \times 2$ matrix.

\section{Irreducible decomposition formulation and ADM variables}
\label{sec:irreducible}

The spectral analysis of the previous section gives a clear, coordinate-independent way of describing the degeneracy structure of the metric Hessian. In the following, we will present an alternative formulation through an irreducible decomposition of the most general kernel element. As expected, the same conclusions concerning the classification of primary constraints are obtained in this formalism. In addition, we present the Hessian, kernel and primary constraints in the ADM formalism.

\subsection{Canonical momenta}

As a starting point, we compute again expressions for the canonical momenta separating explicitly the velocity components on the metric from the nondynamical parts. For this, we write the general expression for the Hessian of Newer GR defined in \eqref{HessP} explicitly in terms of metric components
\begin{equation}
\begin{aligned}
\label{HessianNewerGR}
K^{\alpha\beta\mu\nu}
={}&
c_1 g^{00}\left(g^{\alpha\mu}g^{\beta\nu}+g^{\alpha\nu}g^{\beta\mu}\right)
+2c_3 g^{00} g^{\alpha\beta} g^{\mu\nu}
\\
&+\frac{c_2+c_4}{2}\Big(
g^{0\beta}g^{0\nu}g^{\alpha\mu}
+g^{0\alpha}g^{0\nu}g^{\beta\mu}
+g^{0\beta}g^{0\mu}g^{\alpha\nu}
+g^{0\alpha}g^{0\mu}g^{\beta\nu}
\Big)
\\
&+c_5\left(g^{\alpha\beta}g^{0\mu}g^{0\nu}+g^{\mu\nu}g^{0\alpha}g^{0\beta}\right).
\end{aligned}
\end{equation}
With the components of the Hessian after a $3+1$ split of the spacetime components $\mu=0,i$ computed in \ref{appendix}, and using Eqs. \eqref{eq:momfull} and \eqref{eq:momspatial}, we can find the following expressions for the canonical momenta
\begin{equation}
\begin{split}
&\dfrac{2\pi^{00}}{\sqrt{-g}}   = K^{0000} \dot{g}_{00} + 2 K^{000i} \dot{g}_{0i} + K^{00ij} \dot{g}_{ij} + C^{00},\\
&\dfrac{2 \pi^{0i}}{\sqrt{-g}} = K^{0i00} \dot{g}_{00} + 2 K^{0i0j} \dot{g}_{0j} + K^{0ikl} \dot{g}_{kl} + C^{0i},\\
& \dfrac{2 \pi^{ij}}{\sqrt{-g}} = K^{ij00} \dot{g}_{00} + 2 K^{ij0k} \dot{g}_{0k} + K^{ijkl} \dot{g}_{kl} + C^{ij}.
\end{split}
\label{piKC}
\end{equation}
From now on we will use the shorthand notation $\hat{c} = (c_1+c_2+c_3+c_4+c_5)$, $c_{35}= 2c_3 + c_5$, $c_{124}=c_1 + \dfrac12(c_2 + c_4)$, together with the definitions
\begin{equation}
S_q = q_{000} - 2\beta^{i}q_{00i} + \beta^i \beta^j q_{0ij}, \qquad N_i = q_{00i} - \beta^j q_{0ij},
\end{equation}
\begin{equation}
    T_q = \gamma^{ij} q_{0ij}, \qquad T^{ij}_{q} = \gamma^{ik} \gamma^{jl} q_{0kl},
\end{equation}
and the spatial nonmetricity traces
\begin{equation}
\begin{split}
    Q_i & = g^{\rho\sigma} Q_{i\rho\sigma},\\
    Q_{i}{}^{\mu\nu} & = g^{\mu\rho} g^{\nu\sigma} Q_{i\rho\sigma},\\
    \tilde{Q}^{\mu}_s & = g^{\mu\rho} g^{\lambda k} Q_{k\rho\lambda}.
\end{split}
\end{equation}
With these definitions, the velocity-independent part $B^{\mu\nu}$ written in \eqref{Bmostgeneral} can be expanded as
\begin{equation}
\begin{split}
B^{\mu\nu} & = 2c_1 g^{0k} Q_k{}^{\mu\nu} + c_2\left( g^{\mu k}Q_k{}^{\nu 0} + g^{\nu k}Q_k{}^{\mu 0} \right) + 2c_3 g^{\mu\nu}g^{0k}Q_k \\
& + c_4\left( g^{0\mu}\tilde{Q}_s^\nu + g^{0\nu} \tilde{Q}_s^\mu \right) + \dfrac{c_5}{2} \left( 2g^{\mu\nu}\tilde{Q}_s^0 + g^{0\mu}g^{\nu k}Q_k + g^{0\nu}g^{\mu k}Q_k \right).
\end{split}
\end{equation}
After replacing these expressions in Eqs. \eqref{piKC}, we find
\begin{equation}
\dfrac{2 \pi^{00}}{\sqrt{-g} } = -\dfrac{2}{\alpha^6} \hat{c} \dot{g}_{00} + \dfrac{4}{\alpha^6} \hat{c} \beta^i \dot{g}_{0i} + \left( \dfrac{c_{35}}{\alpha^4}\gamma^{ij} - \dfrac{2\hat{c}}{\alpha^6}\beta^i \beta^j \right) \dot{g}_{ij} + C^{00}
\end{equation}
where
\begin{equation}
C^{00} = B^{00} + \dfrac{2\hat{c}}{\alpha^6} S_q - \dfrac{c_{35}}{\alpha^4} T_{q}.
\end{equation}
On the other hand, $\pi^{0i}$ is:
\begin{equation}
\begin{aligned}
   \dfrac{ 2 \pi^{0i}}{\sqrt{-g}}  &= 2 \frac{\hat{c}}{\alpha^6}(\beta^i\dot{g}_{00} - 2\beta^i \beta^j \dot{g}_{0j}+\beta^i\beta^l\beta^k \dot{g}_{lk}) \\ &+ \frac{c_{124}}{\alpha^4}\Big[2\gamma^{ij}\dot{g}_{0j} -(\beta^l \gamma^{ik} + \beta^k \gamma^{il})\dot{g}_{lk} \Big] - \frac{c_{35}}{\alpha^4}\beta^i \gamma^{lk}\dot{\gamma}_{lk} + C^{0i},
\end{aligned}
\end{equation}
with
\begin{equation}
    C^{0i} = B^{0i} - \dfrac{2\hat{c}}{\alpha^6} \beta^{i} S_q - \dfrac{2 c_{124}}{\alpha^4 } \gamma^{ij} N_{j} + \dfrac{c_{35}}{\alpha^4} \beta^{i} T_{q}.
\end{equation}
Finally, the spatial components of the momenta $\pi^{ij}$ can be written as
\begin{equation}
\begin{aligned}
    \dfrac{2 \pi^{ij}}{\sqrt{-g}} &= \frac{c_{35}}{\alpha^4}\Big[\gamma^{ij}\dot{g}_{00} - 2\beta^k \gamma^{ij}\dot{g}_{0k} + (\gamma^{ij} \beta^l \beta^k + \gamma^{lk}\beta^i \beta^j)\dot{g}_{lk} \Big] \\ &-\frac{2\hat{c}}{\alpha^6}\Big[\beta^i \beta^j \dot{g}_{00} - 2\beta^i \beta^j \beta^k \dot{g}_{0k} + \beta^i\beta^j \beta^l \beta^k \dot{g}_{lk}\Big] \\ &+
    \frac{c_{124}}{\alpha^4}\Big[-2(\beta^i \gamma^{jk} + \beta^j \gamma^{ik})\dot{g}_{0k} + (\gamma^{ik}\beta^j \beta^l + \gamma^{il}\beta^j \beta^k + \gamma^{jk}\beta^i \beta^l + \gamma^{jl}\beta^i \beta^k)\dot{g}_{lk}\Big] \\& -
    \frac{c_1}{\alpha^2}\Big(\gamma^{ik}\gamma^{jl} + \gamma^{il} \gamma^{jk} \Big)\dot{g}_{lk} - \frac{2c_3}{\alpha^2}\gamma^{ij} \gamma^{lk} \dot{g}_{lk} + C^{ij}.
\end{aligned}
\end{equation}
with
\begin{equation}
\begin{split}
C^{ij} & = B^{ij} - \dfrac{c_{35}}{\alpha^4} \gamma^{ij} S_{q} + \dfrac{2\hat{c} }{\alpha^6} \beta^{i} \beta^{j} S_q + \dfrac{ 2c_{124} }{\alpha^{4}}( \beta^{i} \gamma^{jk} + \beta^{j} \gamma^{ik} ) N_k - \dfrac{c_{35}}{\alpha^{4}}\beta^{i} \beta^{j} T_q\\
& + \dfrac{2 c_1}{\alpha^2} T^{ij}_q + \dfrac{2 c_{3}}{\alpha^2} \gamma^{ij} T_q.
\end{split}
\end{equation}
Using the relations \eqref{veloctoadm}, we can also obtain the $\pi^{\mu\nu}$ components fully in terms of the velocities of the ADM variables as 
\begin{equation}
        \frac{2 \pi^{00}}{\sqrt{-g}} =  \frac{4\hat{c}}{\alpha^5}\dot{\alpha} + \frac{c_{35}}{\alpha^4}\gamma^{ij}\dot{\gamma}_{ij} -  \frac{2\hat{c}}{\alpha^6}\beta^i \beta^j \dot{\gamma}_{ij} + C^{00},
\end{equation}
\begin{equation}
        \frac{2\pi^{0i}}{\sqrt{-g}} =  
        -\frac{4\hat{c} \beta^{i} }{\alpha^5} \dot{\alpha} + 2 \frac{c_{124}}{\alpha^4}\dot{\beta}^i  - \Big[\frac{c_{124}}{\alpha^4} (\beta^j \gamma^{ik} + \beta^k \gamma^{ij}) + \frac{c_{35}}{\alpha^4} \beta^i \gamma^{jk} \Big]\dot{\gamma}_{jk} + C^{0i},
\end{equation}
\begin{equation}
\begin{aligned}
    \frac{2\pi^{ij}}{\sqrt{-g}} & = -2\dot{\alpha}\left(\frac{c_{35}}{\alpha^3}\gamma^{ij} -\frac{2\hat{c}}{\alpha^5}\beta^i\beta^j \right) - \dfrac{2c_{124}}{\alpha^4}(\beta^{i} \gamma^{jk} + \beta^{j} \gamma^{ik} ) \dot{\beta}_k  \\ & +\Bigg[\frac{c_{35}}{\alpha^4} \gamma^{kl}\beta^i \beta^j  + \frac{c_{124}}{\alpha^4}(\gamma^{ik}\beta^j \beta^l + \gamma^{il}\beta^j \beta^k + \gamma^{jk}\beta^i \beta^l + \gamma^{jl}\beta^i \beta^k) \\& -
    \frac{c_1}{\alpha^2}\Big(\gamma^{ik}\gamma^{jl} + \gamma^{il} \gamma^{jk} \Big)- \frac{2c_3}{\alpha^2}\gamma^{ij} \gamma^{lk}\Bigg]\dot{\gamma}_{lk} + C^{ij}.
\end{aligned}
\end{equation}
From the previous expressions, we can also obtain the canonical momenta conjugate to the ADM variables as defined in \eqref{lapse00}, \eqref{shift0i} and \eqref{spmomrel} as
\begin{equation}
\overset{\alpha}{\pi} = -\sqrt{\gamma} \left[ \dfrac{4\hat{c}}{\alpha^3} \dot{\alpha} + \left( \dfrac{c_{35}}{\alpha^2}\gamma^{ij} - \dfrac{2\hat{c}}{\alpha^4}\beta^{i} \beta^{j} \right) \dot{\gamma}_{ij} + \alpha^2 C^{00} \right] ,
\end{equation}
\begin{equation}
\overset{\beta}{\pi}{}^{k} = \sqrt{\gamma}\left[ \dfrac{2 c_{124}}{\alpha^{3}} \gamma^{ij} \dot{\beta}_{j} - \left( \dfrac{c_{124}}{\alpha^{3}}(\beta^{j}\gamma^{ik} + \beta^{k}\gamma^{ij}) + \dfrac{2\hat{c}}{\alpha^{5}}\beta^{i} \beta^{j} \beta^{k} \right) \dot{\gamma}_{ij} + \alpha(\beta^{i} C^{00} + C^{0i}) \right],
\end{equation}
and
\begin{equation}
\begin{split}
\overset{\gamma}{\pi}{}^{ij} & = \sqrt{\gamma} \Bigg[ -\dfrac{c_{35}}{\alpha^2} \gamma^{ij} \dot{\alpha} - \dfrac{c_{124}}{\alpha^3} ( \beta^{i}\gamma^{jk} + \beta^{j} \gamma^{ik} ) \dot{\beta}_{k} \\
& +\left( \dfrac{\hat{c}}{\alpha^5} \beta^{i} \beta^{j} \beta^{k} \beta^{l} + \dfrac{c_{124}}{2\alpha^3} ( \gamma^{ik} \beta^{j} \beta^{l} + \gamma^{il} \beta^{j} \beta^{k} + \gamma^{jk} \beta^{i} \beta^{l} + \gamma^{jl} \beta^{i} \beta^{k} ) \right. \\
& \left. - \dfrac{c_1}{2\alpha} ( \gamma^{ik} \gamma^{jl} + \gamma^{il} \gamma^{jk} ) - \dfrac{c_3}{\alpha} \gamma^{ij} \gamma^{kl} \right) \dot{\gamma}_{kl} + \dfrac{\alpha}{2}( C^{ij} - \beta^{i} \beta^{j} C^{00} ) \Bigg].
\end{split}
\end{equation}
As expected, and in comparison to the ADM formulation of GR, the momenta conjugate to the lapse and shift do not vanish identically. All dependence on nondynamical pieces of the connection is contained in the velocity-independent terms $C^{\mu\nu}$, which shift the canonical momenta but do not change the rank of the Hessian, which is our subject of study in the next subsection.

\subsection{Kernel elements}

Here we are interested in obtaining the primary constraints for Newer GR. For this, we must obtain the kernel $X_{\alpha\beta} = X_{\beta\alpha}$ of the Hessian $K^{\alpha\beta\mu\nu}$, which is defined such that
\begin{equation}
K^{\alpha\beta\mu\nu} X_{\mu\nu} = 0
\end{equation}
for all $\mu,\nu$. If $X_{\alpha\beta}$ is an element of the kernel, then in the relation
\begin{equation}
   \dfrac{2\pi^{\alpha\beta}}{\sqrt{-g}}- C^{\alpha\beta} = \mathcal{K}^{\alpha\beta\mu\nu} \dot{g}_{\mu\nu}
\end{equation}
we multiply by $X_{\alpha\beta}$ by the left, obtaining in this way the primary constraints
\begin{equation}
X_{\alpha\beta}\left( \dfrac{2\pi^{\alpha\beta}}{\sqrt{-g}}- C^{\alpha\beta} \right) = X_{\alpha\beta} K^{\alpha\beta\mu\nu} \dot{g}_{\mu\nu} = 0.
\end{equation}
In the way we have written it, the Hessian does not have a kernel for arbitrary values of the $c_i$, and different cases will give a different kernel and consequently different number of primary constraints.

Any arbitrary component of the kernel can be decomposed as
\begin{equation}
X_{\mu\nu} = a n_{\mu} n_{\nu} + b \gamma_{\mu\nu} + 2 n_{(\mu} V_{\nu)} + S_{\mu\nu},
\end{equation}
with $V_{\mu}$ a purely spatial vector, i.e.
\begin{equation}
n^{\mu} V_{\mu} = 0,
\end{equation}
and $S_{\mu\nu}$ a symmetric trace-free tensor such that
\begin{equation}
n^{\mu} S_{\mu\nu} = 0, \ \ g^{\mu\nu} S_{\mu\nu} = \gamma^{\mu\nu} S_{\mu\nu} = 0.
\end{equation}

We can summarize the action of the generic kernel element $X_{\alpha\beta}$ in the Hessian in three components as
\begin{equation}
K^{\alpha\beta\mu\nu} X_{\mu\nu} = K^{\alpha\beta}_{scalar} + K^{\alpha\beta}_{vector}  + K^{\alpha\beta}_{tensor}.
\end{equation}
The contribution of the scalar sector is
\begin{equation}
\begin{aligned}
    K^{\alpha\beta}_{scalar}  
    &= 2 c_1 g^{00}(a n^\alpha n ^\beta + b \gamma^{\alpha\beta}) + 2c_3g^{00}(a n^\alpha n^\beta - a \gamma^{\alpha\beta} + 3 b \gamma^{\alpha\beta} - 3 bn^\alpha n^\beta) \\ &+ c_5 g^{00}(a n^\alpha n^{\beta} - a \gamma^{\alpha\beta} - 3 b n^{\alpha} n^{\beta}) + 2(c_2 + c_4) g^{00}an^{\alpha}n^{\beta}.
\end{aligned}
\end{equation}
Collecting the terms with respect to the scalars $a$ and $b$ we get:
\begin{equation}
\begin{aligned}
    K^{\alpha\beta}_{scalar} &= 2 g^{00}a n^{\alpha}n^{\beta}(c_1 + c_2 + c_3 + c_4 + c_5) - g^{00} a \gamma^{\alpha\beta} (2c_3 + c_5) \\ &- 3g^{00} b n^{\alpha}n^\beta (2c_3 + c_5) + 2g^{00} b\gamma^{\alpha \beta}(c_1 + 3 c_3).
\end{aligned}
\end{equation}
We see that the result can be expressed as the multiplication of a $2\times 2$ matrix spanned by the basis $(n^{\alpha} n^{\beta}, \gamma^{\alpha\beta})$, times the vector $(a,b)$. The multiplication of the transformation matrix with the scalar modes can be written as
\begin{equation}
K^{\alpha\beta}_{scalar} = M =\begin{pmatrix}2 g^{00}(c_1 + c_2 + c_3 + c_4 + c_5) & \hphantom{000} & - 3g^{00} (2c_3 + c_5) \\ {} & {} & {} \\ -g^{00}(2c_3 + c_5) & \hphantom{000} & 2g^{00} (c_1 + 3c_3) \end{pmatrix} \left( \begin{array}{c}
     a \\
    b
\end{array} \right),
\end{equation}
whose determinant, up to a factor of $4(g^{00})^2$ is:
\begin{equation}
\label{eq: scalar}
    \text{det(M)} = c_1^2 + c_1(c_2 +4c_3 +c_4 + c_5) + 3c_3(c_2 + c_4) - \frac{3}{4}c^2_5.
\end{equation}
\ 
Now the action of the Hessian into the vector subkernel gives a simple result:
\begin{equation}\label{eq: vector}
    K^{\alpha\beta}_{vector} = (2c_1 + c_2 + c_4) g^{00} (V^\alpha n^{\beta} + V^{\beta} n^\alpha),
\end{equation}
as well as the action of the Hessian into the tensor modes, which can be simply written as
\begin{equation}\label{eq: tensor}
\begin{aligned}
    K^{\alpha\beta}_{tensor} = 2c_1 g^{00} S^{\alpha\beta}.
\end{aligned}
\end{equation}
This analysis highlights the splitting of the kernel into irreducible parts, showing a simple structure of the action of the Hessian on the kernel. Moreover, the three degeneracy conditions in Eqs. \eqref{eq: scalar}, \eqref{eq: vector}, \eqref{eq: tensor}, for the scalar, vector, and tensor parts, can be written respectively as
\begin{align}
     & c_1^2 + c_1(c_2 +4c_3 +c_4 + c_5) + 3c_3(c_2 + c_4) - \frac{3}{4}c^2_5 = 0,\\
     & 2c_1 + c_2 + c_4 = 0, \\
     &c_1 = 0,
\end{align}
representing the factorization of the Hessian determinant into the three kind of modes. Thus, the ansatz $K^{\alpha\beta\mu\nu}X_{\mu \nu} =0$ equips a full method to determine the possible null directions of the Hessian. The results of this method fully coincide with the spectral analysis developed in Sec.\ref{sec:spectral}. As we will see later, this method also reproduces the degeneracy in the scalar sector that produces one or two scalar primary constraints.

\subsection{Primary constraints}

The tensor sector is controlled by the degeneracy condition $c_1=0$. If this condition is achieved, then every spatial symmetric trace-free tensor $S_{\mu\nu}$ belongs to the kernel of the Hessian. In three spatial dimensions, $S_{\mu\nu}$ has $5$ independent components, hence there are $5$ primary constraints. Defining the expression
\begin{equation}
Y^{\mu\nu} = \dfrac{2\pi^{\mu\nu}}{\sqrt{-g}} - C^{\mu\nu},
\label{YpC}
\end{equation} 
the primary constraints are then
\begin{equation}
S_{\mu\nu} Y^{\mu\nu} = 0,
\end{equation}
for all spatial trace-free $S_{\mu\nu}$. In order to express these constraints in component notation, we define
\begin{equation}
Z^{ij} \equiv Y^{ij} + \beta^{i} Y^{0j} + \beta^{j} Y^{0i} + \beta^{i} \beta^{j} Y^{00}.
\end{equation}
Then the five tensor primary constraints are
\begin{equation}
Z^{ij} - \dfrac13 \gamma^{ij} \gamma_{kl} Z^{kl} \approx 0,
\end{equation}
which can be written in terms of $\pi^{\mu\nu}$ and $C^{\mu\nu}$ by replacing back $Y^{\mu\nu}$ as in Eq. \eqref{YpC}. 

In the vector sector, every spatial vector $V_{\mu}$ belongs to the kernel if the condition $2c_1 + c_2 + c_4 = 0$ is satisfied. The three independent components of this arbitrary spatial vector give rise to $3$ primary constraints given by
\begin{equation}
2 n_{\mu} V_{\nu} Y^{\mu \nu} \approx 0.
\end{equation}
In coordinate components, this can be written as
\begin{equation}
g^{00}\left( \dfrac{2\pi^{0i}}{\sqrt{-g}} - C^{0i} \right) - g^{0i}\left( \dfrac{2\pi^{00}}{\sqrt{-g}} - C^{00} \right) \approx 0.
\label{Pi0iconstr}
\end{equation}

The scalar sector is two-dimensional, spanned by $n_{\mu} n_{\nu}$ and $\gamma_{\mu\nu}$, becoming degenerate when the parameters vanish the combination \eqref{eq: scalar}. If the scalar block has rank $1$, then the kernel is one-dimensional, and there is exactly $1$ scalar primary constraint. Let $(a,b)\neq (0,0)$ be a kernel vector of the scalar matrix
\begin{equation}
\left( 
\begin{array}{cc}
   2 \hat{c}  & -3 c_{35}  \\
    - (2c_3+c_5) & 2(c_1+3c_3) 
\end{array}
\right) \left(
\begin{array}{c}
     a \\
     b 
\end{array}
\right) = 0.
\end{equation}
Then, the scalar primary constraint is
\begin{equation}
a n_{\mu} n_{\nu} Y^{\mu\nu} + b \gamma_{\mu\nu} Y^{\mu\nu} \approx 0.
\end{equation}
It could be chosen a convenient generic kernel vector such as $a=2(c_1+3c_3)$, $b=2c_3 + c_5$.

The scalar sector gives two constraints only if the full $2\times 2$ scalar block vanishes, not merely if its determinant vanishes. Thus, one needs simultaneously, that
\begin{equation}
    \begin{split}
       & 2c_3+c_5 = 0\\
       & c_1 + 3c_3 = 0,\\
       & \hat{c} = 0.
    \end{split}
\end{equation}
In this case, both scalar directions belong to the kernel:
\begin{equation}
X_{\mu\nu} = a n_{\mu} n_{\nu}, \qquad X_{\mu\nu} = b \gamma_{\mu\nu}.
\end{equation}
Therefore, the two scalar primary constraints are
\begin{equation}
n_{\mu} n_{\nu} Y^{\mu\nu} \approx 0, \qquad \gamma_{\mu\nu} Y^{\mu\nu} \approx 0,
\end{equation}
which in components, up to a factor $\alpha^2$, are written as
\begin{equation}
\dfrac{2 \pi^{00}}{\sqrt{-g}} - C^{00} \approx 0
\end{equation}
and
\begin{equation}
\gamma_{ij}\left( \dfrac{2 \pi^{ij}}{\sqrt{-g}} - C^{ij}  \right) + 2 \beta_{i} \left( \dfrac{2\pi^{0i}}{\sqrt{-g}} - C^{0i}  \right) + \beta_{i} \beta^{i} \left( \dfrac{2\pi^{00}}{\sqrt{-g}} - C^{00} \right) \approx 0,
\end{equation}
respectively. Notice that these scalar constraints have not been previously found in the literature.

\section{Review of the possible models}
\label{sec:models}

We have seen that the Hessian of Newer GR decomposes into irreducible tensor, vector and scalar blocks of dimensions $5$, $3$ and $2$, respectively. The corresponding primary constraints are therefore naturally classified according to the degeneracies of these three sectors. The tensor sector is degenerate when $c_1=0$, and then contributes with five primary constraints. The vector sector is degenerate when
\begin{equation}
2c_1 + c_2 + c_4=0,
\end{equation}
and contributes with the three primary constraints \eqref{Pi0iconstr}. Finally, the scalar sector is governed by a two-dimensional Hessian block. Generically, its degeneracy gives one scalar primary constraint, from the vanishing of the quadratic equation \eqref{quadreq}. Only on a smaller locus does the scalar block vanish completely, giving two scalar primary constraints. This happens when
\begin{equation}
4c_1 = -3(c_2 +c_4)= -12 c_3 = 6c_5,
\label{two-scalar}
\end{equation}
or, equivalently,
\begin{equation}
c_1=-3c_3,
\qquad
c_2+c_4 =4c_3,
\qquad
c_5=-2c_3 .
\end{equation}
The tensor, vector and scalar degeneracies cannot be combined arbitrarily. Naively, since the tensor and vector sectors may each be either degenerate or nondegenerate, while the scalar sector may have zero, one or two null directions, one might expect $2 \times 2 \times 3 = 12$ possible patterns. However, two of these possibilities are algebraically excluded.

Indeed, suppose first that the tensor condition $c_1=0$ is imposed together with the two-scalar condition \eqref{two-scalar}. Then $c_1 = -3c_3 = 0$, hence $c_3=0$. It follows that $c_2+c_4=0$, $c_5=0$. But then the vector condition $2c_1 + c_2 + c_4$ is automatically satisfied. Therefore, a sector with two scalar and five tensor primary constraints cannot occur. Whenever the tensor kernel and the two-dimensional scalar kernel coexist, the vector kernel is necessarily present as well. Similarly, suppose that the vector condition $2c_1 + c_2 + c_4 = 0$ is imposed together with the two-scalar condition. Using $c_1=-3c_3$ and $c_2+c_4 = 4 c_3$, one obtains $c_3 = 0$, therefore $c_1=c_2+c_4=c_5=0$, so the tensor condition is automatically satisfied. Therefore, the only option is that the tensor, vector and two-dimensional scalar kernel coexist, which is a case that contains ten primary constraints. This is a model of fully degenerate kinetic sector, or more precisely, a model with no unconstrained momenta at all.

The remaining mixed cases have simple algebraic characterizations. Combining the vector degeneracy with a one-dimensional scalar kernel reduces the scalar condition \eqref{quadreq} to
\begin{equation}
\label{grsc}
c_1^2 + c_1 (2c_3 - c_5) +\frac34 c_5^2=0.
\end{equation}

In particular, the STEGR model belongs to this class. This sector is not exhausted by STEGR itself. Indeed, one only needs that the change in three parameters $c_1$, $c_3$, and $c_5$ does not violate the equation (\ref{grsc}). If $c_1$ did have a non-vanishing variation, it can be compensated by a change in $c_2+c_4$ for preserving the vector constraint.

At the same time, if we change only the $c_3$ coefficient relative to STEGR, it corresponds to a family of Newer GR models known as one-parameter newer GR \cite{Bello-Morales:2024vqk}. It removes the scalar constraint, and therefore belongs to another class. It can be viewed as a unimodular-like modification of STEGR \cite{Golovnev:2024owe}.

Combining instead the tensor degeneracy with a one-dimensional scalar kernel reduces the conditions on the parameters to
\begin{equation}
4c_3 (c_2 + c_4) = c_5^2,
\end{equation}
which reproduces the sector previously identified as Case IV in Ref. \cite{Dambrosio:2020wbi}. Finally, the vector and tensor degeneracies taken together imply $c_1=c_2+c_4=0$. This sector can be further combined with one scalar constraint by imposing $c_5=0$, with $c_3 \neq 0$ corresponding to a one-dimensional scalar kernel. Otherwise, the scalar block becomes fully degenerate and the theory reaches the maximally degenerate case with ten primary constraints.

The complete classification of all possible models is summarized in Table \ref{tab:primary-constraint-combinations}. The table also indicates the relation with the sector decomposition of Ref. \cite{Dambrosio:2020wbi}. The irreducible tensor-vector-scalar decomposition makes explicit a distinction that is not fully resolved in that reference: the scalar sector may have either a one- or two-dimensional kernel. In particular, the case with two primary constraints but no tensor or vector constraints is not isolated as an independent sector in that reference. Likewise, the fully degenerate sector is naturally understood here as the simultaneous presence of 10 primary constraints.
\begin{table}[h]
\centering
\begin{tabular}{c c c c c c}
\hline
Case & Tensor& Vector & Scalar & NPC & Relation to  \cite{Dambrosio:2020wbi} \\
\hline
0 & no  & no  & no              & $0$  & Sector $0$ \\
1 & no  & no  & $1$ scalar       & $1$  & Part of Sector I / scalar-deg. sector \\
2 & no  & no  & $2$ scalars      & $2$  & Not present \\
3 & no  & yes & no              & $3$  & Sector II \\
4 & no  & yes & $1$ scalar       & $4$  & Sector V (includes GR) \\
5 & yes & no  & no              & $5$  & Sector III \\
6 & yes & no  & $1$ scalar       & $6$  & Sector IV \\
7 & yes & yes & no              & $8$  & Sector VI \\
8 & yes & yes & $1$ scalar       & $9$  & Sector VII \\
9 & yes & yes & $2$ scalars      & $10$ & Sector VIII / fully deg. case \\
\hline
\end{tabular}
\caption{Possible combinations of primary constraints from the tensor, vector and scalar sectors of the Hessian. The tensor sector contributes $5$ constraints, the vector sector contributes $3$ constraints, and the scalar sector contributes either $1$ or $2$ constraints. NPC denotes number of primary constraints. Note that Sector VIII in \protect\cite{Dambrosio:2020wbi} (Case 9 in our work) it is not resolved as a sector with two scalar primary constraints, therefore the equivalence is not full.}
\label{tab:primary-constraint-combinations}
\end{table}
Thus, apart from the completely nondegenerate case, there are nine nontrivial primary-constraint patterns, with total number of primary constraints $1, 2, 3, 4, 5, 6, 8, 9$ and $10$, where the absence of a seven-constraint sector reflects precisely the algebraic incompatibilities described above.

\section{Covariant primary constraints}
\label{sec:cpc}

Now we discuss several approaches to obtain the covariant primary constraints appearing in any gravity based in symmetric teleparallel geometry. A first approach for considering the connection in the Hamiltonian formalism would be to treat $\Gamma^{\rho}{}_{\mu\nu}$ as a fundamental variable, with conjugate canonical momenta $\Pi_{\rho}{}^{\mu\nu}$ defined as in \eqref{momgG}. Since the connection is symmetric in the torsionless sector, the corresponding fundamental Poisson bracket might be written as
\begin{equation}
\{ \Gamma^{\rho}{}_{\alpha\beta}(\textbf{x}) , \Pi_{\sigma}{}^{\mu\nu}(\textbf{y}) \} = \frac12 \delta^{\rho}_{\sigma} \left(\delta^{\mu}_{\alpha} \delta^{\nu}_{\beta} + \delta^{\nu}_{\alpha} \delta^{\mu}_{\beta} \right) \delta^{(3)}(\textbf{x}-\textbf{y}).
\end{equation}
However, in such a formulation the conditions of vanishing curvature and torsion do not come naturally and would have to be imposed by Lagrange multipliers. This enlarges the constraint system and complicates the analysis.

Schematically,  $\dot\Gamma$ enters the Lagrangian density only linearly in the Lagrange multiplier approach. Consequently, the canonical Hamiltonian does not depend on $\Pi_{\rho}{}^{\mu\nu}$, and these momenta appear in the total Hamiltonian only through primary-constraint terms. Then, preservation of the secondary constraint $\Gamma_{[,]}=0$, which imposes vanishing torsion, entails the requirement that the antisymmetric part of the corresponding Lagrange multipliers vanish. In turn, this means that only the symmetric part of $\Pi_{\rho}{}^{\mu\nu}$ actually enters any of the Hamiltonian equations.

Alternatively, one may try to define canonical momenta directly for the scalar fields $\xi^{a}$ determining the symmetric teleparallel connection. Since the connection depends on second derivatives of $\xi^{a}$, this would require an Ostrogradsky procedure. Introducing the auxiliary variable $\Xi = \dot{\xi}^{a}$, one would define two momenta
\begin{equation}
\Pi_a = \dfrac{\partial \mathcal{L}}{\partial \dot{\xi}^{a}} - \partial_t \left( \dfrac{\partial \mathcal{L}}{ \partial \ddot{\xi}^{a} } \right) = \dfrac{\partial \mathcal{L}}{\partial {\Xi}^{a}} - \partial_t \left( \dfrac{\partial \mathcal{L}}{ \partial \dot{\Xi}^{a} } \right),
\label{Pi_a1}
\end{equation}
and
\begin{equation}
P_{a} = \dfrac{\partial \mathcal{L}}{\partial \ddot{\xi}^{a}} = \dfrac{\partial \mathcal{L}}{\partial \dot{\Xi}^{a}}.
\label{Pi_a2}
\end{equation}
Considering the pairs of canonical variables $(\xi^{a}, \Pi_{b})$ and $(\Xi^{a}, P_{b})$, we define the Poisson brackets as \cite{Woodard:2015zca}
\begin{equation}
    \{\xi^{a}(\textbf{x}), \Pi_{b}(\textbf{y}) \} = \delta^{a}_{b} \delta^{(3)}(\textbf{x}-\textbf{y}), \qquad \{ \Xi^{a}(\textbf{x}), P_{b}(\textbf{y}) \} = \delta^{a}_{b} \delta^{(3)}(\textbf{x}-\textbf{y}).
\end{equation}
It is worth noticing the unusual nature of this formulation, since the $\xi^{a}$ play the role of coordinate fields as well as dynamical variables. Moreover, the Ostrogradsky construction in \eqref{Pi_a1} and \eqref{Pi_a2} suggests that the momenta will appear linearly in the Hamiltonian, which as a result will give a Hamiltonian unbounded from below and be prone to ghosts. However, recently in Ref. \cite{ErrastiDiez:2024hfq} it was discussed the possibility of stability in theories with unbounded conserved quantities.

In order to avoid these potential problems, we explore an alternative route towards a covariant formulation in the Hamiltonian analysis, which consists in reducing the order of the derivatives in the connection. In \cite{Blixt:2023kyr}, a tetrad-like approach to the symmetric teleparallel connection was proposed, by defining
\begin{equation}
e^a{}_\mu=\frac{\partial \xi^a}{\partial x^\mu},
\label{covtetr}
\end{equation}
Let $E_{a}{}^{\mu}$ denote the inverse matrix
\begin{equation}
E_{a}{}^{\mu} e^{a}{}_{\nu} = \delta^{\mu}_{\nu}, \qquad E_{a}{}^{\mu} e^{b}{}_{\mu} = \delta^{b}_{a}.
\end{equation}
The symmetric teleparallel connection is then written as
\begin{equation}
\Gamma^{\rho}{}_{\mu\nu} = E_{a}{}^{\rho} \partial_{\nu} e^{a}{}_{\mu}.
\end{equation}
The integrability condition
\begin{equation}
\partial_\mu e^a{}_\nu-\partial_\nu e^a{}_\mu = 0
\end{equation}
ensures that $e^{a}{}_{\mu}$ is locally of the form $\partial_{\mu} \xi^{a}$, and also enforces the torsionless condition for the connection above. Since this condition is no longer automatic when $e^{a}{}_{\mu}$ is treated as an independent variable, it must be imposed via Lagrange multipliers $ \lambda_{a}{}^{\mu\nu}$. Therefore, we add to the Lagrangian density the term
\begin{equation}
    \lambda_{a}{}^{\mu\nu} ( \partial_\mu e^a{}_\nu - \partial_\nu e^a{}_\mu).
\end{equation}
We also introduce the inverse $E_{a\mu} = g_{\mu\nu} E_{a}{}^{\nu}$ with lowered spacetime index. The symmetric teleparallel connection can thus be written as 
\begin{equation}
\Gamma_{\rho\mu\nu} = g_{\rho \sigma} \frac{\partial x^\sigma}{\partial \xi^a}\frac{\partial^2 \xi^a}{\partial x^\mu\partial x^\nu} = E_{a\rho} \partial_\nu e^{a}{}_\mu.
\end{equation}
As a consequence, the nonmetricity tensor can be written as
\begin{equation}
Q_{\rho\mu\nu} = \partial_{\rho} g_{\mu\nu} - E_{a\nu} \partial_{\rho} e^{a}{}_{\mu} - E_{a \mu} \partial_{\rho} e^{a}{}_\nu.
\label{nonm_tetr}
\end{equation}
Notice that $E_{a\mu}$ is not the same object as $e^{a}{}_{\mu}$. The latter is not a metric tetrad, and therefore cannot be used to lower the spacetime index of the inverse matrix.
We now define the momenta conjugate to $e^{a}{}_{\sigma}$ as
\begin{equation}
P^{a\sigma} = \dfrac{\partial \mathcal{L}}{\partial \partial_0 e^{a}{}_{\sigma}} = \frac{\partial\mathcal L}{\partial Q_{\rho\mu\nu}}
\frac{\partial Q_{\rho\mu\nu}}
{\partial(\partial_0 e^a{}_\sigma)}.
\end{equation}
The inverse $E_{a}{}^{\mu}$ depends algebraically on $e^{a}{}_{\mu}$, but not on its derivatives, and therefore it does not contribute to the derivative with respect to $\partial_0 e^{a}{}_{\sigma}$. 
We have that
\begin{equation}
\frac{\partial\mathcal L}{\partial Q_{\rho\mu\nu}} = \sqrt{-g} \mathcal P^{\rho\mu\nu},
\end{equation}
and the variation of the nonmetricity tensor in terms of the tetrads \eqref{nonm_tetr} gives the following
\begin{equation}
\frac{\partial Q_{\rho\mu\nu}}{\partial(\partial_0 e^a{}_\sigma)} = - \delta^0_\rho (E_{a \nu}  \delta^\sigma_\mu + E_{a \mu} \delta^\sigma_\nu).
\end{equation}
The new momenta are then
\begin{equation}
P_{a}{}^{\sigma} = -2 E_{a\nu} \pi^{\sigma \nu} + \lambda_{a}{}^{0\sigma} - \lambda_{a}{}^{\sigma 0},
\end{equation}
which, if choosing the multiplier to be antisymmetric $\lambda_{a}{}^{\mu\nu} = - \lambda_{a}{}^{\nu\mu}$, gives rise to the following covariant primary constraints
\begin{equation}
\chi_{a}{}^{\sigma} = P_{a}{}^{\sigma} + 2 E_{a\nu} \pi^{\sigma\nu} - 2 \lambda_{a}{}^{0\sigma} \approx 0.
\end{equation}
These constraints are the covariant analogue of the covariant primary constraints appearing in the tetrad formulation of metric teleparallel gravity, found in Ref. \cite{Golovnev:2021omn}. They express the fact that, once the connection is written in first-order pure-gauge form, the momenta conjugate to the connection variables are not independent from the metric momenta. As it was hypothesized in \cite{Golovnev:2021omn}, we expect these constraints to be first-class, although a rigorous proof would require to write the Hamiltonian and prove that these constraints commute with the remaining primary and secondary constraints. This issue will be considered for future work.

\section{Conclusions}
\label{sec:concl}

We have analyzed the structure of primary constraints of the most general parity-even Newer General Relativity. This amounts to carrying out the first step of the Dirac-Bergmann algorithm: computing the canonical momenta, identifying the degeneracies of the velocity-momenta map, and determining the corresponding primary constraints. Since these degeneracies are controlled by the kinetic part of the Lagrangian, the resulting classification depends only on the Hessian computed with respect to the metric velocities.

The main result of our analysis is that the Hessian naturally decomposes into tensor, vector and scalar sectors, each with its own degeneracy condition and corresponding primary constraints. The tensor sector contributes five primary constraints when it becomes degenerate, while the vector sector contributes three. The scalar sector has a richer structure: it is governed by a two-dimensional block, whose degeneracy generically produces one scalar primary constraint, but on a smaller parameter locus the whole scalar block vanishes and gives two independent scalar primary constraints. This distinction between one and two scalar primary constraints is one of the central differences of the present work compared to previous ones. 

Combining the tensor, vector and scalar sectors leads to a complete classification of the possible primary-constraint patterns. Apart from the nondegenerate case, the allowed total numbers of primary constraints are $1,2,3,4,5,6,8,9$ and $10$. The absence of a seven-constraint sector follows from algebraic relations among the degeneracy conditions: in particular, whenever the two-dimensional scalar kernel is combined with either the tensor or vector kernel, the remaining sector is forced to become degenerate as well, leading to the maximally degenerate case with ten primary constraints.

In terms of the irreducible decomposition, the construction is simple in its vector and tensor sectors, and the corresponding constraints were correctly identified in previous works \cite{Dambrosio:2020wbi, Bajardi:2024qbi}. Note that the paper \cite{Bajardi:2024qbi} is more general because it deals with general teleparallel models, though the scalar sector is not properly analyzed there. At the same time, Ref. \cite{Dambrosio:2020wbi} claims an incorrect value of the Hessian determinant, with complicated functions of metric components depending on the numbering of spatial dimensions. As a result, it does not present the full structure of the scalar constraints.

We have shown that the most general condition for constraints in the scalar sector, in any spacetime dimension, is vanishing of $2\times 2$ matrix determinant, that is the equation (\ref{quadreq}) in $3+1$ dimensions. We have seen that the cases of scalar constraints in the paper \cite{Dambrosio:2020wbi} are actually correct, but they correspond to only some special cases of the more general equation \eqref{quadreq}.

In addition to what is described above, we discussed how the connection may be incorporated in a covariant Hamiltonian treatment. Treating the connection components as completely independent variables requires additional constraints enforcing vanishing curvature and torsion. If we propose the scalar fields $\xi^a$ as fundamental variables, we face risks concerning Ostrogradsky instabilities. Alternatively, introducing first-order variables $e^{a}{}_{\mu} = \partial_{\mu} \xi^{a}$ avoids the direct appearance of second derivatives of the $\xi^{a}$, but it requires the enforcement of vanishing torsion through Lagrange multipliers. In this approach, we were able to find the primary constraints from the covariant formulation, which is a novel result from this work. These constraints provide the natural starting point for a fully covariant Hamiltonian analysis of symmetric teleparallel theories beyond the coincident gauge.

Although our motivation comes from symmetric teleparallel gravity, the primary-constraint analysis performed here is controlled only by the kinetic Hessian of the quadratic nonmetricity action. In this sense, the result may also be read more broadly as a classification of the degeneracies of a globally Lorentz-covariant quadratic kinetic structure for a symmetric rank-two tensor. The factor $\sqrt{-g}$ in the integration measure does not affect the rank of the Hessian, and therefore does not change the primary-constraint conditions. The tensor-vector-scalar decomposition also extends straightforwardly to arbitrary spacetime dimension, with the only dimension-dependent part being the explicit form of the scalar $2 \times 2$ block.

The present work does not yet determine the final number of propagating degrees of freedom. For that, one must continue the Dirac-Bergmann algorithm: construct the primary Hamiltonian, impose time preservation of primary constraints, determine the secondary constraints, and classify the full constraint set into first- and second-class. Despite the fact that already the linearized gravity limit of Newer GR possesses complicated features \cite{Golovnev:2024owe}, the present work is the necessary starting point in the Hamiltonian analysis of the symmetric teleparallel gravity models. A natural extension to this analysis is to study the 3+1 decomposition of the equations of motion and well-posedness for viable classes of Newer GR models, following recent progress in the 3+1 formulation in TEGR \cite{Pati:2022nwi,Peshkov:2022cbi,Cheng:2026hvn} and STEGR \cite{Capozziello:2021pcg,Guzman:2023oyl}. We hope to present this analysis in future works.

\section*{Acknowledgments}
M.J.G. has been supported by the Estonian Research Council grant PSG910 ``Theoretical frameworks for numerical modified gravity''. The computations in this paper have been partially assisted with Cadabra 2.0 \cite{Peeters:2006kp,Peeters:2007wn,Peeters:2018dyg}.

\section*{ORCID}

\noindent Carmen Ferrara - \url{https://orcid.org/0000-0001-5623-2706}

\noindent Alexey Golovnev - \url{https://orcid.org/0000-0002-7821-1516}

\noindent Mar\'ia Jos\'e Guzm\'an - \url{https://orcid.org/0000-0001-7468-2647}

\appendix

\section{Intermediate computations}
\label{appendix}

In this appendix we compile different formulas that have been used in the results presented in Sec. \ref{sec:irreducible}.
The explicit independent components of the Hessian \eqref{HessianNewerGR} read
\begin{equation}
\begin{split}
K^{0000} & =2(c_1+c_2+c_3+c_4+c_5)(g^{00})^3, \\
K^{000i} & =2(c_1+c_2+c_3+c_4+c_5)(g^{00})^2 g^{0i}, \\
K^{00ij} & = \left(2c_1+2c_2+2c_4+c_5\right)g^{00}g^{0i}g^{0j}
+ \left(2c_3+c_5\right)(g^{00})^2 g^{ij}, \\
K^{0i0j} & = \left(c_1+\frac{3}{2}(c_2+c_4)+2c_3+2c_5\right)g^{00}g^{0i}g^{0j}
+ \left(c_1+\frac{1}{2}(c_2+c_4)\right)(g^{00})^2 g^{ij}, \\
K^{0ijk} & = \left(c_1+\frac{1}{2}(c_2+c_4)\right)g^{00}\left(g^{0j}g^{ik}+g^{0k}g^{ij}\right)
+ \\
& \left(2c_3+c_5\right)g^{00}g^{0i}g^{jk} + \left(c_2+c_4+c_5\right)g^{0i}g^{0j}g^{0k},\\
K^{ijkl}
& = 
c_1 g^{00}\left(g^{ik}g^{jl}+g^{il}g^{jk}\right)
+2c_3 g^{00} g^{ij} g^{kl}
\\
&+\frac{c_2+c_4}{2}\left(
g^{0j}g^{0l}g^{ik}
+g^{0i}g^{0l}g^{jk}
+g^{0j}g^{0k}g^{il}
+g^{0i}g^{0k}g^{jl}
\right)
\\
&+c_5\left(g^{ij}g^{0k}g^{0l}+g^{kl}g^{0i}g^{0j}\right).
\end{split}
\end{equation}
Due to the following symmetries of the Hessian
\begin{equation}
K^{\alpha\beta\mu\nu}=K^{\beta\alpha\mu\nu} =K^{\alpha\beta\nu\mu}=K^{\beta\alpha\nu\mu} =K^{\mu\nu\alpha\beta},
\end{equation}
the previous components of the Hessian are dependent on each other
\begin{equation}
K^{0ijk}=K^{0ikj},
\qquad
K^{ijkl}=K^{jikl}=K^{ijlk}=K^{klij}.
\end{equation}
In ADM coordinates, the components of the Hessian become
\begin{equation}
\begin{split}
K^{0000} & =-\frac{2(c_1+c_2+c_3+c_4+c_5)}{\alpha^6},\\
K^{000i} & =\frac{2(c_1+c_2+c_3+c_4+c_5)}{\alpha^6}\beta^i, \\
K^{00ij} & = \frac{2c_3+c_5}{\alpha^4}\gamma^{ij}
-\frac{2(c_1+c_2+c_3+c_4+c_5)}{\alpha^6}\beta^i \beta^j,\\
K^{0i0j} & = \frac{c_1+\frac{1}{2}(c_2+c_4)}{\alpha^4}\gamma^{ij}
-\frac{2(c_1+c_2+c_3+c_4+c_5)}{\alpha^6}\beta^i \beta^j,\\
K^{0ijk} & = -\frac{c_1+\frac{1}{2}(c_2+c_4)}{\alpha^4}
\left(\beta^j \gamma^{ik}+\beta^k \gamma^{ij}\right)
-\frac{2c_3+c_5}{\alpha^4}\beta^i \gamma^{jk}
+\frac{2(c_1+c_2+c_3+c_4+c_5)}{\alpha^6}\beta^i \beta^j \beta^k, \\
K^{ijkl} & = 
-\frac{c_1}{\alpha^2}\left(\gamma^{ik}\gamma^{jl}+\gamma^{il}\gamma^{jk}\right)
-\frac{2c_3}{\alpha^2}\gamma^{ij}\gamma^{kl}
\\
&+\frac{c_1+\frac{1}{2}(c_2+c_4)}{\alpha^4}
\left(
\gamma^{ik}\beta^j \beta^l
+\gamma^{il}\beta^j \beta^k
+\gamma^{jk}\beta^i \beta^l
+\gamma^{jl}\beta^i \beta^k
\right)
\\
&+\frac{2c_3+c_5}{\alpha^4}
\left(
\gamma^{ij}\beta^k \beta^l
+\gamma^{kl}\beta^i \beta^j
\right)
-\frac{2(c_1+c_2+c_3+c_4+c_5)}{\alpha^6}\beta^i \beta^j \beta^k \beta^l.
\end{split}
\end{equation}

\bibliographystyle{unsrt}
\bibliography{bibl.bib}

\end{document}